\documentclass[twocolumn,aps,prl, superscriptaddress]{revtex4-1}
\usepackage{amsmath, lettrine}
\usepackage{placeins}
\usepackage{booktabs}
\usepackage{color}
\usepackage{cleveref, array}
\usepackage{subfigure}
\usepackage{graphicx}

\begin{document}

\title{Inefficient magnetic-field amplification in supersonic laser-plasma turbulence}

\author{A.F.A.~Bott}
\affiliation{Department of Physics, University of Oxford, Parks Road, Oxford OX1 3PU, UK}
\affiliation{Department of Astrophysical Sciences, University of Princeton, 4 Ivy Ln, Princeton, NJ 08544, USA}

\author{L.~Chen}
\affiliation{Department of Physics, University of Oxford, Parks Road, Oxford OX1 3PU, UK}

\author{G. Boutoux}
\affiliation{CEA-DAM, DIF, F-91297 Arpajon, FRANCE}

\author{T. Caillaud}
\affiliation{CEA-DAM, DIF, F-91297 Arpajon, FRANCE}

\author{A. Duval}
\affiliation{CEA-DAM, DIF, F-91297 Arpajon, FRANCE}

\author{M.~Koenig}
\affiliation{LULI, CNRS, CEA, Ecole Polytechnique, 
UPMC, Univ Paris 06: Sorbonne Universites,
Institut Polytechnique de Paris, F-91128 Palaiseau cedex, France}
\affiliation{Graduate School of Engineering, Osaka University, Suita, Osaka 565-0871, Japan}

\author{B. Khiar}
\affiliation{Department of Astronomy and Astrophysics, University of Chicago, 5640 S. Ellis Ave, Chicago, IL 60637, USA}

\author{I. Lantu\'ejoul}
\affiliation{CEA-DAM, DIF, F-91297 Arpajon, FRANCE}

\author{L. Le-Deroff}
\affiliation{CEA, DAM, CESTA, F-33114 Le Barp, France}

\author{B. Reville}
\affiliation{Max-Planck-Institut f\"ur Kernphysik, Postfach 10 39 80, 69029 Heidelberg, Germany}

\author{R. Rosch}
\affiliation{CEA-DAM, DIF, F-91297 Arpajon, FRANCE}

\author{D. Ryu}
\affiliation{Department of Physics, School of Natural Sciences, UNIST, Ulsan 44919, Korea}

\author{C. Spindloe}
\affiliation{Central Laser Facility, STFC Rutherford Appleton Laboratory, Chilton, Didcot OX11 0XQ, UK}

\author{B. Vauzour}
\affiliation{CEA-DAM, DIF, F-91297 Arpajon, FRANCE}

\author{B. Villette}
\affiliation{CEA-DAM, DIF, F-91297 Arpajon, FRANCE}

\author{A.A.~Schekochihin}
\affiliation{Department of Physics, University of Oxford, Parks Road, Oxford OX1 3PU, UK}
\affiliation{Merton College, Merton St, Oxford OX1 4JD, UK}

\author{D.Q.~Lamb}
\affiliation{Department of Astronomy and Astrophysics, University of Chicago, 5640 S. Ellis Ave, Chicago, IL 60637, USA}

\author{P. Tzeferacos}
\affiliation{Department of Physics, University of Oxford, Parks Road, Oxford OX1 3PU, UK}
\affiliation{Department of Astronomy and Astrophysics, University of Chicago, 5640 S. Ellis Ave, Chicago, IL 60637, USA}
\affiliation{Department of Physics and Astronomy, University of Rochester, 206 Bausch \& Lomb Hall, Rochester, NY 14627, USA}
\affiliation{Laboratory for Laser Energetics, University of Rochester, 250 E. River Rd, Rochester, New York 14623, USA}

\author{G. Gregori}
\affiliation{Department of Physics, University of Oxford, Parks Road, Oxford OX1 3PU, UK}

\author{A. Casner}
\affiliation{CEA-DAM, DIF, F-91297 Arpajon, FRANCE}
\affiliation{Universit\'e de Bordeaux-CNRS-CEA, CELIA, UMR 5107, F-33405 Talence, France}

\begin{abstract}

We report a laser-plasma experiment that was carried out at the LMJ-PETAL facility and realized the first magnetized, turbulent, supersonic plasma with a large magnetic Reynolds number ($\mathrm{Rm} \approx 45$) in the laboratory. Initial seed magnetic fields were amplified, but only moderately so, and did not become dynamically significant. A notable absence of magnetic energy at scales smaller than the outer scale of the turbulent cascade was also observed. Our results support the notion that moderately supersonic, low-magnetic-Prandtl-number plasma turbulence is inefficient at amplifying magnetic fields.

    
\end{abstract}

\maketitle 

Understanding the kinematics and dynamics of magnetic fields in supersonic plasma turbulence is 
a challenge that has both its own intrinsic merit and important astrophysical 
applications. Compared to the thorough characterizations of     
supersonic turbulent boundary layers arising from aerofoils~\citep{SSR94,L94},
which have been tested by numerous 
computer simulations~\citep[e.g.,][]{PM07,PB11}
and wind-tunnel experiments~\citep{SD94,RB06}, 
current theories of magnetized, supersonic, boundary-free plasma turbulence have a much weaker 
empirical foundation. At the same time, the wide range of physical 
processes that can arise in such a system promises an exceptionally 
rich collection of complex phenomena for study.  In the astrophysical context, magnetic 
fields are believed to play a significant role in the turbulent, supersonic dynamics of the 
interstellar medium (ISM); 
understanding the complex interactions between the fields, shocks and vortices 
present in such an environment is a necessary 
component of a comprehensive picture of the ISM, encompassing important topics 
such as star formation~\citep{PVP96,LK04,KF19,G20}. 
{Magnetized, moderately supersonic plasma turbulence is also  
thought to emerge in solar and stellar convection zones~\citep{C90,NSA90}}.

One key question concerning the relationship between magnetic fields and 
supersonic plasma turbulence in the ISM is how the fields attain their observed 
dynamical strengths. The equivalent question in subsonic plasma turbulence has 
been studied in greater depth, mostly within the framework of magnetohydrodynamics (MHD). 
Analytical theory~\citep{K67,ZRMS84,KA92,BC04,S02}, simulations~\citep{MFP81,SCTM04,HBD04,S07,SBSW20}, and recent 
experiments~\citep{M14,M15,T18,B20} give a consistent picture, showing that 
chaotic bulk motions of plasma (with characteristic scale $L$ and velocity $u_{\rm rms}$) can amplify any small seed magnetic 
field initally present in the plasma provided the \textit{magnetic Reynolds number} 
$\mathrm{Rm} \equiv u_{\rm rms} L/\eta$ is greater than a certain critical value $\mathrm{Rm}_{c}$ (here, $\eta$ is the plasma resistivity). {This critical value is usually significantly larger than unity~\citep{R19}. } 
For $1 \ll \mathrm{Rm} \lesssim \mathrm{Rm}_c$, the magnitude $\delta B$ of the magnetic field post amplification is related to the magnitude $B_0$ of  
the initial seed field via $\delta B \sim \mathrm{Rm}^{1/2} B_0$~\citep{S07}. However, if $\mathrm{Rm} > 
\mathrm{Rm}_c$, magnetic-field amplification of seed fields proceeds unabated until 
the magnetic-energy density of the amplified field reaches equipartition with 
the kinetic-energy density of the stochastic motions responsible for the amplification; 
this field-amplification mechanism is known as the \textit{fluctuation dynamo}. {Another important parameter for magnetic-field amplification is the magnetic Prandtl number $\mathrm{Pm} \equiv \mathrm{Rm}/\mathrm{Re}$ (where $\mathrm{Re}$ is the fluid Reynolds number): dynamo is less efficient for $\mathrm{Pm} \ll 1$ than for $\mathrm{Pm} \gtrsim 1$~\citep{BC04,S07}.}

Compared to the subsonic case, there exist far fewer studies of magnetic-field 
amplification in supersonic plasma turbulence. Numerical studies of supersonic MHD turbulence~\citep{HBM04,F16,FSBS14,PR19}
 indicate that the fluctuation dynamo is still capable of operating. 
The efficacy of the mechanism, both in terms of the 
characteristic growth rates of magnetic fields and saturated magnetic/kinetic 
energy ratios, is a function of the turbulent Mach number $\mathrm{Ma}_{\rm turb} \equiv u_{\rm rms}/c_s$ 
(where $c_s$ is the plasma's sound speed): it is less effective for $\mathrm{Ma}_{\rm turb} \gtrsim 1$ 
than for $\mathrm{Ma}_{\rm turb} \ll 1$. 
{Physically, this has been attributed to a number of factors: reduced energy available to the solenoidal stretching motions necessary for dynamo action on account of some of the driving kinetic energy flux being directed towards compressive motions, irrespective of the driving mechanism~\citep{PWP98,SNP02,HBM04,MB06,F16}; a steepened turbulent velocity spectrum~\citep{S12}; 
and enhanced dissipation of magnetic fields in shocks~\citep{PN99}. }
In the laboratory, there has only been one previous experiment 
that successfully realized boundary-free, supersonic plasma turbulence~\citep{W19}; 
however, $\mathrm{Rm}$ achieved in that experiment was much smaller than unity, 
prohibiting significant magnetic-field amplification. 

\begin{figure}
  \centering
\includegraphics[width=\linewidth]{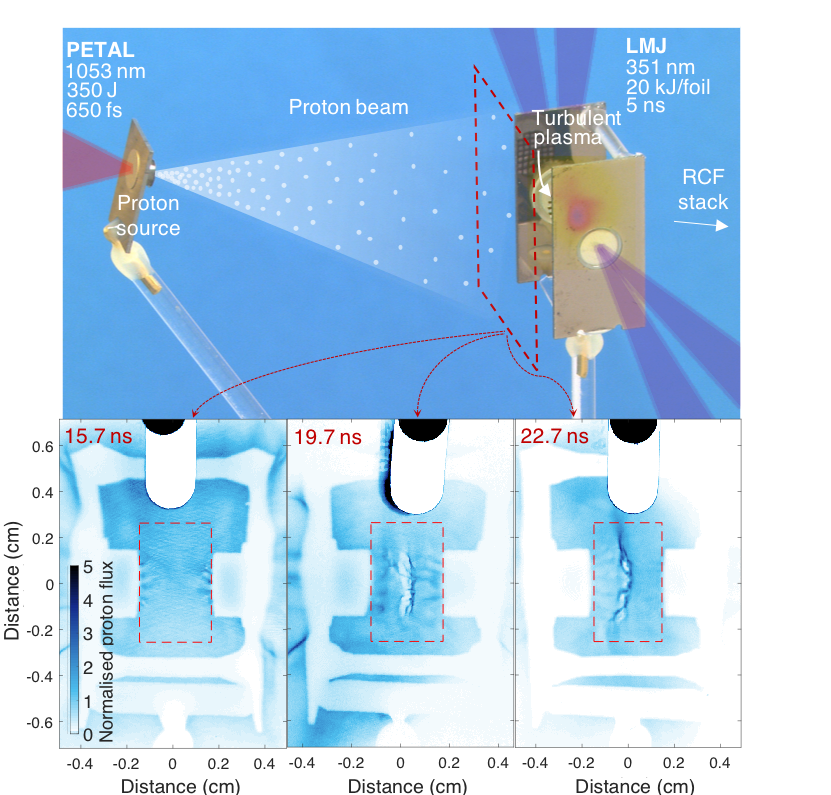}
\caption{\textit{Experimental set-up}. Upper panel: annotated photograph of one of the targets used in our experiment. 
The parameters of both the LMJ and PETAL beams are indicated on the photograph. 
Aluminium foils (separated by 8 mm) are irradiated by the LMJ drive-beam lasers; they have a 3 mm diameter and 25 
$\mu$m thickness. An annular CH washer (230 $\mu$m thick, 3 mm diameter, 400 $\mu$m hole) 
is placed over the foil to aid jet collimation. 
The grids (located 2 mm away from each foil, on the opposite side to the LMJ beams) are made of polyamide, 
have a thickness of 250 $\mu$m, and square holes (side length 300 $\mu$m) separated 
by 100 $\mu$m rods. The main target is rendered partially transparent, in order to show the location of the turbulent plasma 
(the yellow-purple region). The proton source is a 50 $\mu$m gold foil, and is protected
from pre-plasma and return currents by an aluminium polycarbonate shield. It is located 3 cm from the turbulent plasma's centre. The RCF stack 
used to detect the protons after they pass through the plasma is placed 10 cm away on the opposite side, leading to a $\times 4.3$ magnification. All length scales are shown with this magnification factor removed, i.e., on the plasma's scale.
Lower panels: 8.5 MeV proton images (obtained from different experimental shots) at 15.7 ns (left), 19.7 ns (middle) and 22.7 ns (right) after the initiation of the LMJ drive beams. 
The proton flux normalisation is defined relative to the mean of the regions enclosed by red-dashed lines in each image.
There was a 30\% drop in delivered LMJ beam energy on both foils for the 15.7 ns experimental shot, and on one foil for the 22.7 ns shot; however, due to inefficient beam-energy absorption in the foil, we do not
 believe that our results are significantly affected by this~\citep{SM}.} 
\label{fig:expsetup}
\end{figure}

In this paper, we report a new experiment that managed to create 
supersonic, high-$\mathrm{Rm}$ plasma turbulence for the first time in the laboratory. 
The experiment was performed on the Laser Megajoule (LMJ) facility in Bordeaux~\citep{C15}. 
The platform employed for the experiment is illustrated in Figure 
\ref{fig:expsetup}.
Similarly to previous laser-plasma experiments investigating the fluctuation dynamo in 
subsonic plasma, which were carried out on the Omega Laser Facility~\citep{T18,B20}, a turbulent plasma was created by colliding inhomogeneous, asymmetric, 
counter-propagating rear-side-blow-off plasma jets. Spatial inhomogeneity is 
introduced by placing grids in the paths of each jet prior to their collision; 
the jet asymmetry follows directly from using asymmetric grids. In order to 
reach the supersonic regime,
three major design modifications to the
previous Omega experiments were introduced. The thickness of the foils irradiated 
by the LMJ drive beams was reduced, and the beam energy per foil increased 
fourfold: both changes led to increased initial jet velocities. In addition, aluminium rather than plastic foils were 
used in the experiment; the resulting enhancement in radiative cooling reduced the plasma's temperature
both before and after jet collision. Both modifications were anticipated to 
increase $\mathrm{Ma}_{\rm turb}$, a claim supported by three-dimensional, 
three-temperature radiation-MHD simulations performed concurrently to the 
experiment using the FLASH code~\citep{F00,KT20}. 

The primary diagnostic on 
the experiment, {CRACC (\textit{Cassette de Radiographie au Centre Chambre})~\citep{L18}}, provides time-resolved proton imaging~\citep{M04}, which was used to measure magnetic fields {and the electron number density }in the plasma, as well as to determine the characteristic velocities of 
the initial jets. 
The proton imaging beam was generated by irradiating a gold foil with the high-intensity PETAL 
beam (see Figure \ref{fig:expsetup})~\citep{B17};
via the target normal sheath acceleration (TNSA) mechanism~\citep{W01}, this irradiation results in a highly directed 
proton beam with a thermal ($\sim 3$ MeV temperature) spectrum. The beam 
passed through the plasma generated by the LMJ drive beams, and subsequently was 
detected using a calibrated radiochromic film (RCF) stack~\citep{L18}.
The RCF stack was designed in such a way 
that protons with distinct energies were detected in separate layers of RCF 
($\sim 0.5$ MeV energy resolution); this allowed for time-resolved measurements on each experimental shot, because
 slower beam protons passed through the plasma at later times than faster 
 ones. The $\Delta t_{p} \approx$ 300 ps time delay between the fastest and slowest detected protons (8.5 MeV vs. 4.7 MeV) was 
 too small to capture the full dynamical evolution of the plasma turbulence; to 
 capture this evolution, we repeated our experiment, but with three different relative 
 offsets between the LMJ and PETAL beams. The resulting proton images (for the
8.5 MeV protons) are shown in Figure \ref{fig:expsetup} (see~\citep{SM} for further information about the analysis of the RCF stack).
 
Detailed quantitative information about the magnetic fields 
present in the turbulent plasma can be obtained by analyzing the proton images. 
The theoretical basis for such analysis comes from the proton beam's high 
velocity and low density compared to that of the plasma with which it interacts prior 
to reaching the RCF stack; inhomogeneites in the detected proton flux can therefore be 
attributed to the action on the beam protons of the Lorentz forces arising from spatially varying magnetic fields present in the 
plasma~\citep{K12}. Collisionless beam instabilities and deflections due to electric fields in this experiment 
have a negligible effect on the proton beam. This being the case, recent work~\citep{AGGS16}
has shown that the two components of the path-integrated magnetic field 
that are perpendicular to the proton beam's direction can be reconstructed 
directly from these inhomogeneities, provided the proton beam, on account of its non-uniform distortion, does not self-intersect 
before reaching the detector. Further 
information on how this analysis was performed is given in~\citep{SM}. 
 
\begin{figure}
  \centering
\includegraphics[width=\linewidth]{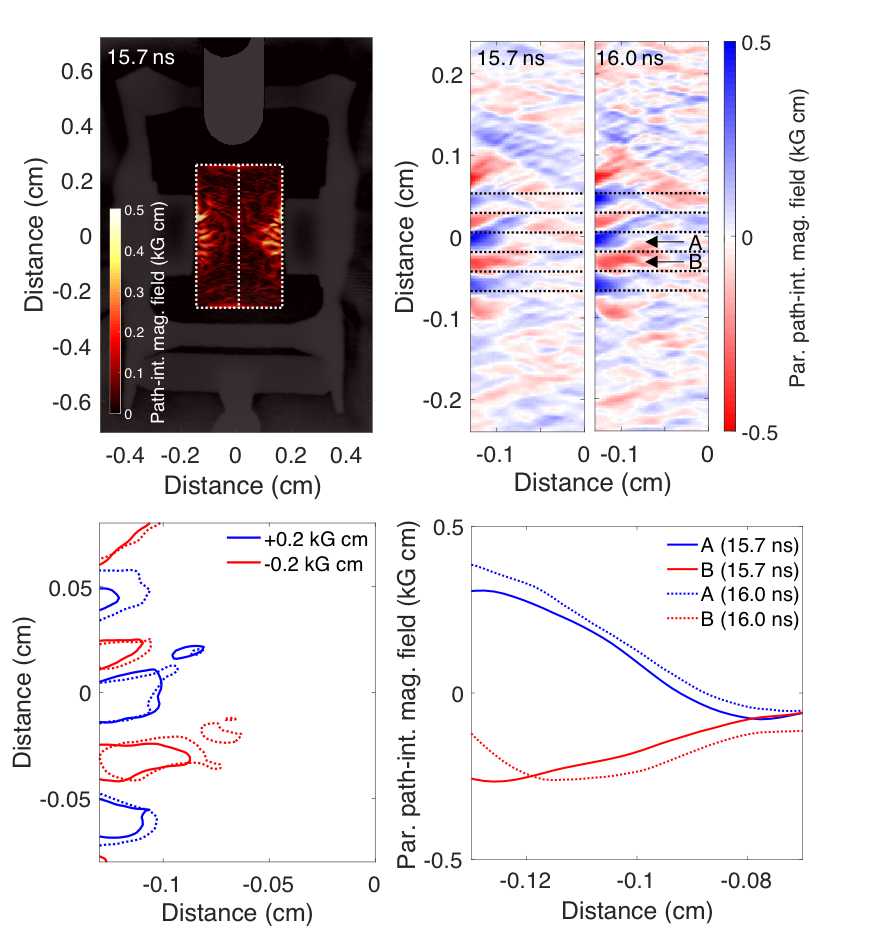}
\caption{\textit{Jet-velocity measurement}. Top left: magnitude of perpendicular path-integrated magnetic field 
reconstructed from the region denoted in the 15.7 ns proton image shown in Figure \ref{fig:expsetup}. The procedure used to extract this 
quantity is described in~\citep{SM} (see also~\citep{AGGS16}). Top right: axial component of the path-integrated magnetic field determined at 15.7 ns and 16.0 ns 
by analyzing 8.5 MeV and 4.7 MeV proton images respectively. Bottom left: $\pm 0.2$ kG cm contour plots of the axial path-integrated magnetic field components at 15.7 ns (solid) and 16.0 ns (dotted). 
Bottom right: lineouts of axial path-integrated magnetic-field component, calculated from the regions `A' and `B' shown above.} 
\label{fig:jetvel}
\end{figure}

The path-integrated field reconstructed from the 8.5 MeV proton image of the supersonic 
plasma jets prior to their collision is given in Figure \ref{fig:jetvel} (top 
left); its time evolution can be used to determine the velocity $u_{\rm jet}$ of the jets.  
The dominant component of the path-integrated field (characteristic magnitude $\sim 0.4$ kG cm) 
inside the main bulk of each jet is aligned with the jets' predominant direction of motion, and oscillates strongly in the  
direction normal to it.
To extract the velocity of the left-hand jet, we compare the 
path-integrated field recovered from 4.7 and 8.5 MeV proton images generated on 
the same experimental shot (see Figure \ref{fig:jetvel}, top right), corresponding to imaging times of 
15.7 and 16.0 ns, respectively. While the 
morphology of both images is very similar, the characteristic extent in the parallel direction of the 
oscillatory path-integrated field structure is slightly greater 
at 16.0 ns; we illustrate this  
qualitatively using contour plots of the path-integrated field (Figure \ref{fig:jetvel}, bottom 
left). We attribute this finding to the motion of the magnetic fields inside the jet: these fields are frozen into the bulk flow provided the jet's magnetic Reynolds number $\mathrm{Rm}_{\rm jet} \equiv u_{\rm jet} L/\eta$ (where $L = 0.04$ cm is the grid periodicity) just after its interaction with the grid satisfies $\mathrm{Rm}_{\rm jet} \gg 1$ (an assumption supported by theoretical expectations concerning the initial jet properties -- see~\citep{SM}).
The mean jet velocity $u_{\rm jet}$ is obtained as follows: calculate average lineouts for five 
different regions (which are depicted in Figure \ref{fig:jetvel}, top right) for the 
path-integrated fields measured at each time (two sample lineouts are shown in Figure \ref{fig:jetvel}, bottom right); determine the mean 
spatial offset $\Delta x_p$ between each temporal pair of lineouts; then estimate $u_{\rm jet}$ via $u_{\rm jet} \approx \Delta x_p/\Delta t_p$. We 
find $u_{\rm jet} = 290 \pm 40 \, \mathrm{km/s}$; this value is consistent 
with a heuristic estimate determined from the known temporal delay between the LMJ drive-beam 
pulse's midpoint and the jet collision time, and the 4 mm distance from each foil to the target's centre. 

 Once collision between the jets has occurred, X-ray imaging from related experiments on other laser facilities~\citep{T18,B20} indicate that a 
 turbulent plasma with higher characteristic temperatures and densities quickly coalesces; this 
 coincides with a burst of self-emitted X-rays. The spectrum of these X-rays was 
 measured in our experiment using the DMX diagnostic~\citep{B01,C16}. DMX is an absolutely calibrated, time-resolved broadband spectrometer with high temporal resolution ($\simeq 100 \, \mathrm{ps}$). The brightness temperature of the 10 lower energy channels (taking into account an X-ray emissive area corresponding to the collision zone)  allow for the turbulent plasma's temperature to be 
 extracted: $T \approx 100 \, \mathrm{eV}$. 

Given our previous measurement of $u_{\rm jet}$, the characteristic turbulent velocity $u_{\rm turb}$ in the interaction-region plasma can be estimated as follows. X-ray measurements from previous experiments~\citep{B20} and FLASH simulations~\citep{KT20} indicate that, while the jet velocities are close to being uniform transversely, the density of either of the
plasma flows is much larger at transverse spatial positions 
coincident with the locations of the grid holes through which that flow has passed than the density at the analogous position in the opposing flow. When the two plasma flows collide, conservation of momentum therefore dictates that the flow velocity in these transverse spatial locations will be close to the higher-density plasma flow's incident velocity. 
Taking into consideration the two-dimensional periodic reversals in the flow 
direction, and assuming that this flow profile is efficiently randomized by nonlinear interactions and/or Kelvin-Helmholtz instabilities, we conclude that $u_{\rm turb} \approx u_{\rm jet}/{\sqrt{2}} \approx 200 \, \mathrm{km/s}$. The sound speed in the plasma is $c_s = \sqrt{\gamma (Z+1) T/m_i} \approx 80 \,
\mathrm{km/s}$, where $\gamma$ is the adiabatic index, $Z$ the plasma's ionization state, and $m_i$ the ion mass. Therefore, the turbulent Mach number is $\mathrm{Ma}_{\rm turb} \approx 2.5$, so the turbulence is supersonic. 

\begin{figure}
  \centering
\includegraphics[width=\linewidth]{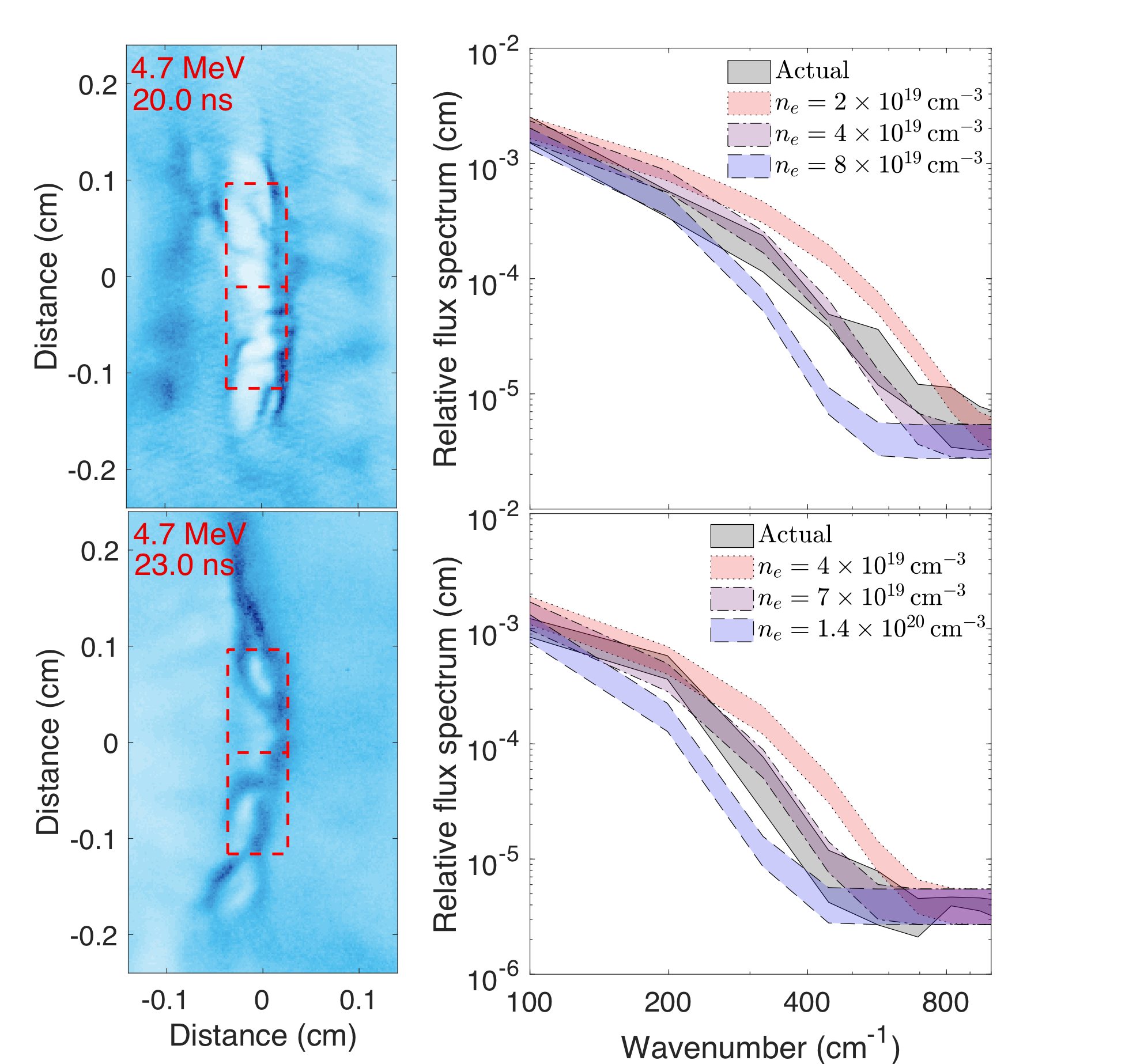}
\caption{\textit{Measurement of the electron number-density of the interaction-region plasma}. Top left: samples from 4.7 MeV proton image at $t = 20.0$ ns (same normalization as for Figure \ref{fig:expsetup}). Top right: 
spectrum of relative 4.7 MeV proton flux (black), as well as the predicted spectra determined by our model at three different electron number densities~\citep{SM}. The mean and the error for each spectrum are calculated by combining the individual results from the regions demarcated by the dashed red lines.} 
\label{fig:denstemp}
\end{figure}

The characteristic electron number density $n_e$ of the interaction-region plasma was determined by quantifying the effect of collisional scattering on the resolution of the sharp, large-amplitude proton-flux inhomogeneities (`caustics') present in the 4.7 MeV proton images. 
In the absence of collisions of the proton beam (and other finite-resolution effects), the Fourier spectrum of caustics is known to follow a characteristic power law $\propto k^{-1}$~\citep{AGGS16}. However,
 Figure \ref{fig:denstemp} shows that, for $k \gtrsim 250$, the measured spectrum of the 4.7-MeV-proton flux inhomogeneities (at both $t = 20.0$ ns and $t = 23.0$ ns) is much steeper. 
Assuming that collisional scattering is the dominant process that limits the resolution of the proton images, the electron density can be estimated using known relations between the characteristic collisional-scattering angle and 
 the image resolution~\citep{SM}. 
We find $n_e \approx 4$--$7 \times 10^{19} \, \mathrm{cm}^{-3}$, a value which is consistent with measurements from related experiments~\citep{T18,B20}.
 
Using all this information, the viscosity and resistivity of the plasma -- and thus the fluid and magnetic Reynolds numbers -- are determined via 
known expressions for transport coefficients in an collisional, aluminium plasma~\citep{SM}. We find that $\mathrm{Re} \approx 10^{6}$, a Reynolds number
which is certainly large enough to allow for the formation of a developed 
turbulent cascade. The magnetic Reynolds number is also significantly larger than unity, but is much smaller than $\mathrm{Re}$:  $\mathrm{Rm} \approx 
45$, {so $\mathrm{Pm} \approx 4 \times 10^{-5}$ }. The turnover time of the turbulence is $\tau_{L} \approx 2$ ns, which is short compared to the lifetime of the interaction-region plasma. 

\begin{figure}
  \centering
\includegraphics[width=\linewidth]{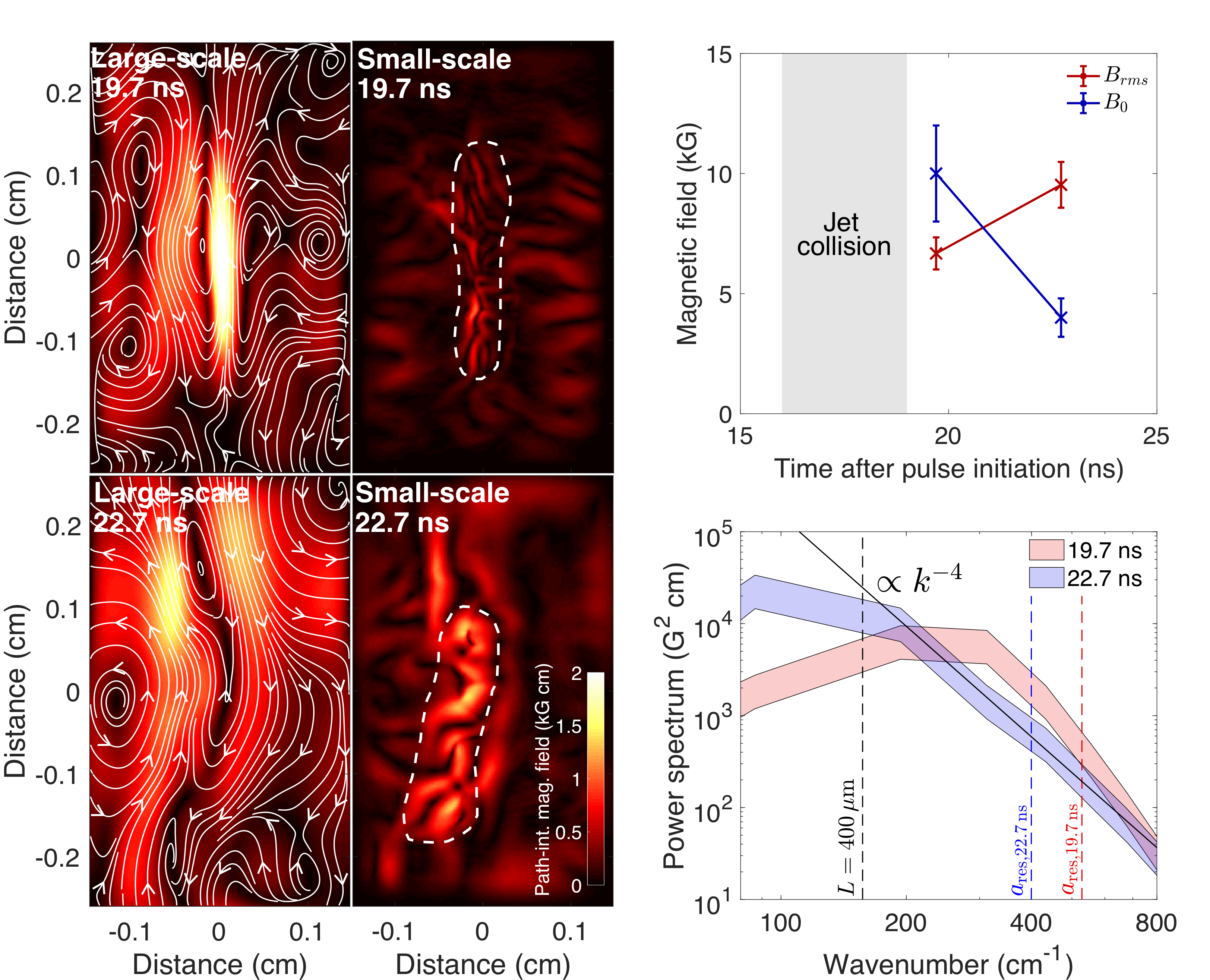}
\caption{\textit{Magnetic-field measurements}. Left: magnitude of perpendicular path-integrated seed (large-scale) magnetic fields at 19.7 ns (top) and 22.7 ns (bottom). The (two-dimensional) streamlines of the perpendicular field are also depicted. 
Middle: magnitude of perpendicular path-integrated stochastic (small-scale) magnetic fields. Top right: evolution of seed (blue) and stochastic (red) magnetic field over time. Bottom right: magnetic-energy spectra calculated in demarcated regions 
from maps of path-integrated stochastic field components. The anticipated resolution limit on our spectra imposed by collisional scattering of the 8.5-MeV proton beam at both times is also shown. 
} 
\label{fig:magfield}
\end{figure}

The path-integrated magnetic-field maps extracted from 8.5 MeV proton images after the jet collision 
allow us to characterize both the seed fields initially present in the  
interaction-region plasma, and the stochastic field structures arising from the 
interaction of those seed fields with the supersonic plasma turbulence. The seed 
fields, which are generated at the laser spots by the Biermann battery~\citep{GRM15} and subsequently advected into the interaction region, have a characteristic transverse scale 
comparable to that of the interaction region ($\ell_{n\perp} \approx 0.25 \, \mathrm{cm}$), while the correlation length of the stochastic fields 
is significantly smaller ($\ell_B \approx 150 \, \mu \mathrm{m}$). 
We take advantage of this scale separation to extract 
distinct path-integrated field maps for the seed and stochastic magnetic fields 
in the experiment (see Figure \ref{fig:magfield}).  
The extraction procedure for the large- and small-scale  path-integrated magnetic fields is explained in~\citep{SM}. 

We estimate the characteristic magnitude $B_{0}$ of the seed magnetic fields via 
a simple relation~\citep{K12}: $B_0 \approx 10 [B_{\rm path,0}(\mathrm{kG \, cm})/2 \, \mathrm{kG \, cm}][\ell_{n\perp} (\mathrm{cm})/0.25 \, \mathrm{cm}]^{-1} \, \mathrm{kG}$, where $B_{\rm path,0}$ 
is the characteristic magnitude of the path-integrated seed magnetic field. The 
field magnitude obtained just after the collision has occurred ($B_0 \approx 10 \, \mathrm{kG}$ at $t = 19.7$ ns after the LMJ drive beams are initiated) 
is consistent with related experiments~\citep{B20}. Over one turnover time later than the collision ($t = 22.7$ ns), 
the seed fields decay considerably ($B_0 \approx 4$ kG), which can be
attributed to their dilution due to the interaction-region plasma's expansion, and turbulent diffusion. 

The stochastic component of the 
magnetic field is characterised by its magnetic-energy spectrum $E_B(k)$, which describes the 
distribution of the magnetic energy amongst different length scales. We determine $E_B(k)$ from the path-integrated map
of the stochastic magnetic field by assuming statistical 
homogeneity and isotropy; under these assumptions, it can be 
shown that $E_B(k) = k E_{\rm path} (k)/4 \pi^2 \ell_{n\perp}$, where $E_{\rm path}(k)$ 
is the one-dimensional spectrum of the path-integrated field~\citep{AGGS16}. The 
root-mean-square of the stochastic field, $\delta B_{\rm rms}$, can then be calculated directly from $E_B(k)$ 
as $\delta B_{\rm rms} = [8 \pi \int_0^{\infty} \mathrm{d}k \, E_B(k)]^{1/2}$. We find 
that at $t = 19.7$ ns, $\delta B_{\rm rms} \approx 6 \, \mathrm{kG}$, before subsequently attaining magnitudes 
comparable in strength to the initial seed fields ($\delta B_{\rm rms} \approx 10 \, \mathrm{kG}$ at $t = 22.7$ 
ns -- see Figure \ref{fig:magfield}). The magnetic-energy spectra at 
both times have steep power-law tails $E_B(k) \propto k^{-4}$, with the 
spectral peaks at wavenumber $k_{\rm peak} \approx 2 \pi/L$, where $L$ is the grid periodicity. 

Our measurements suggest that amplification of the magnetic fields by 
the supersonic turbulence is quite limited, in spite of $\mathrm{Rm}$ being significantly greater than unity. 
The peak amplification factor of the seed field (at $t = 22$ ns) is $\sim 2.5$ (significantly below the ${\rm Rm}^{1/2} \approx 6.3$-times growth expected by simulations for incompressible flows if below critical~\citep{S07}); this 
is similar to the amplification seen in previous subsonic plasma turbulence experiments at much lower $\mathrm{Rm}$~\citep{M15,SM}. 
The magnetic-kinetic energy ratio is $E_{\rm mag}/E_{\rm kin} \approx  10^{-4}$, a value well below
the saturation values found in simulations of MHD turbulent supersonic dynamos or in subsonic dynamo experiments~\citep{T18,FSBS14,SM}.   
This suggests that we did not reach the dynamo regime in our experiment, in turn providing a lower bound on 
$\mathrm{Rm}_c$ {for $\mathrm{Pm} \ll 1$}. 

{In summary, our results are broadly consistent with the expectation that magnetic-field amplification is less efficient in supersonic, low-$\mathrm{Pm}$ turbulence, as compared to moderate-$\mathrm{Pm}$ subsonic turbulence. In spite of this inefficiency, we believe that creating a laser-plasma turbulent dynamo in the supersonic regime in future experiments is feasible. FLASH simulations of the LMJ experiment, which (by assuming more efficient laser-target energy coupling than was attained in the experiment) realized characteristic kinetic and thermal energies ${\sim}3$-$4$ times greater than we report here, achieve $\mathrm{Rm} \approx 750$, and also show the key signatures of dynamo action~\citep{KT20}. This simulation finding suggests that exploring the transition to the dynamo regime in the laboratory is possible: a tantalizing prospect.} 

\acknowledgements
The research leading to these results received funding from the Engineering and Physical Sciences Research Council, grant numbers EP/M022331/1, EP/N014472/1 and EP/RO34737/1, and the U.S. Department of Energy (DOE) National Nuclear Security Administration (NNSA) under Field Work Proposal No. 57789 to ANL, Subcontract No. 536203 with LANL, Subcontract B632670 with LLNL, and grants No. DE-NA0002724, DE-NA0003605, and DE-NA0003934 to the University of Chicago, DE-NA0003539 to the Massachusetts Institute of Technology, and Cooperative Agreement DE-NA0003856 to the LLE, University of Rochester. We acknowledge support from the U.S. DOE Office of Science Fusion Energy Sciences under grant No. DE-SC0016566 and the National Science Foundation under grants No. PHY-1619573, PHY-2033925, and AST-1908551, the France And Chicago Collaborating in The Sciences (FACCTS) Program, and grants 2016R1A5A1013277 and 2017R1A2A1A05071429 of the National Research Foundation of Korea.
Awards of computer time were provided by the U.S. Department of Energy Innovative and Novel Computational Impact on Theory and Experiment (INCITE) and ASCR Leadership Computing Challenge (ALCC) programs. Simulations supporting this research used resources of the Argonne Leadership Computing Facility at ANL, which is supported by the Office of Science of the U.S. DOE under contract DE-AC02-06CH11357. 
Support from AWE plc. and the Science and Technology Facilities Council of the United Kingdom is also acknowledged. 
The PETAL laser was designed and constructed by CEA under the financial auspices of the Conseil Regional d’Aquitaine, the French Ministry of Research, and the European Union. The CRACC diagnostic was designed and commissioned on the LMJ-PETAL facility as a result of the PETAL+ project coordinated by University of Bordeaux and funded by the French Agence Nationale de la Recherche under grant ANR-10-EQPX-42-01. The LMJ-PETAL experiment presented in this article was supported by Association Lasers et Plasmas and by CEA. 
{We are grateful to the staff of CEA-DAM and LMJ for making the experiment possible.}

\end{document}

\title[Supersonic plasma turbulence on LMJ]{SUPPLEMENTAL MATERIAL\\
{\Large Inefficient magnetic-field amplification in supersonic laser-plasma turbulence}\\
Bott et al.}

\maketitle

\section{Effect of suppressed laser performance on results}

As stated in the main text, in the shot during which the 15.7 ns proton image 
was obtained, a 30\% drop in the laser energy (14 kJ per foil vs. 20 kJ per foil) delivered to both foils by LMJ was 
recorded, due to a technical failure; an identical drop on one foil was recorded in the shot during which the 22.7 ns proton 
image was obtained. Here, we explain why we believe that this change in delivered 
drive-beam energy does not significantly affect our results: specifically, that for our 
experimental parameters, the initial velocity of the rear-side-blow-off plasma jet is only weakly sensitive to the drive beam energy.  

This claim can be justified as follows. Given a laser focal-spot diameter of $300 \, \mu$m, a pulse length of 5 ns, and 
a total beam energy of 20 kJ per foil, the laser intensity $I_L$ is given by $I_{L} \approx 5.6 \times 10^{15} \, \mathrm{W} \, 
\mathrm{cm}^{-2}$. For a laser wavelength $\lambda_L = 351$ nm and laser intensities $I_L \sim 10^{13}$--$10^{15}  \, \mathrm{W} \, 
\mathrm{cm}^{-2}$, a simple physical argument~\citep{FF790}
suggests that the ablation pressure $P_{\rm ab}$ 
approximately scales with $I_L$ via $P_{\rm ab} \propto I_L^{2/3}$ -- a scaling which 
has been (approximately) verified experimentally (see, for example,~\citep{F11}). 
In turn, simple scaling arguments show that the characteristic velocity $u_{\rm jet}(s)$ of the rear-side-blow-off plasma jet 
at displacement $s$ from its initial position is related to $P_{\rm ab}$ by $u_{\rm jet}(s) \sim \sqrt{P_{\rm ab} s/\delta_{\rm t} \rho_{\rm 
t}}$, where $\delta_{\rm t}$ is the thickness of the target foil onto which the laser 
is incident, and $\rho_{\rm t}$ is the foil density. This implies that $u_{\rm jet}(s) \propto 
I_L^{1/3}$. In short, a $\sim 30 \%$ reduction in the laser intensity only 
results in a modest ($\sim 10\%$) reduction in the initial jet velocity. In fact, for the characteristic intensities employed in this experiment 
-- for our laser parameters, $[I_L (\mathrm{W} \, \mathrm{cm}^{-2})/10^{15} \, \mathrm{W} \, \mathrm{cm}^{-2}] [\lambda_L(\mu \mathrm{m})/1 \, \mu \mathrm{m}]^{2} \approx 0.7$ 
--  the dependence of the ablation pressure on the laser intensity is likely 
even weaker, on account of various physical phenomena (nonlinear inverse 
Bremsstrahlung
and resonant plasma instabilities) 
that reduce absorption 
efficiency as the intensity is increased further~\citep{M81}.  
In consequence, the error induced in our results by the drop in delivered laser energy is less than 10\%; this is 
smaller than the reported error of our measurements. 

{There is also qualitative experimental evidence that the suppressed laser performance does not affect our results: specifically, the central location of proton-flux inhomogeneities in the ${\sim}$22.7 ns proton image (see Figure \ref{fig:protimags_shot3}). As stated earlier, the drive on one of the foils was 30\% lower than on the other foil in the experimental shot during which this data was collected; if this drop in energy affected the initial velocity of one of the plasma jets, it would be anticipated that the collision of the two jets would happen off-center. However, the mean positions of the proton-flux inhomogeneities in the 19.7-ns and 22.7-ns proton images were, in fact, very similar.}

\section{Additional information on proton-image analysis}

\subsection{Extracting proton flux measurements from RCF film pack}

We convert the raw RCF images obtained in our experiment to (normalised) proton 
flux images using the following procedure. The RCF (HD-V2) films are electronically scanned, and 
then converted into 300-dots-per-inch RGB TIFF files. The red colour channel of 
these files is transformed to an optical density image, and then to a 
deposited energy (MeV/mm$^2$) image using an appropriate calibration for our 
chosen RCF film type. A GEANT-4 simulation~\citep{A03}
was then carried out, in order to determine the relationship between the energy 
deposited in each piece of film by a given proton, and that proton's energy. 
Curves showing this relationship (`the RCF response function') for the six distinct pieces of RCF film used in  
our experiment are shown in Supplementary Figure~\ref{fig:RCFresponse}. 
\begin{figure}
  \centering
\includegraphics[width=0.6\linewidth]{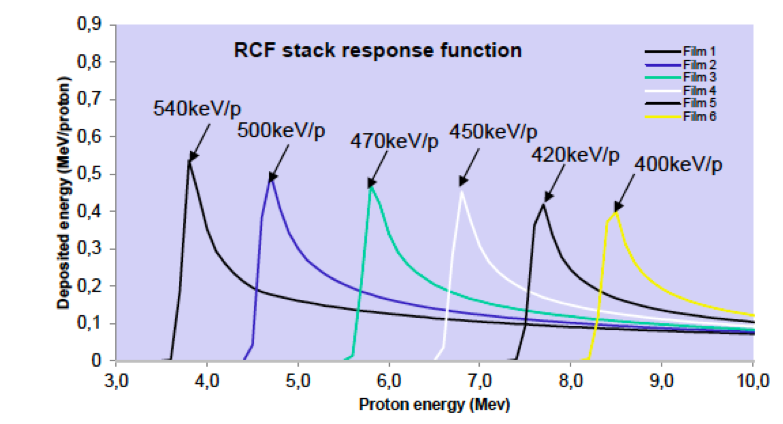}
\caption{\textit{RCF stack response functions}. Relationship between the energy deposited in a given film layer by an imaging proton, and that proton's energy.}
\label{fig:RCFresponse}
\end{figure}
Finally, an estimate of the proton 
flux is made by assuming that the majority of the energy deposited in each 
film comes from the protons that deposit the maximum energy associated with 
that particular piece of film. Table \ref{table:GEANT4} shows this energy for 
each film layer. 
\begin{table}[htbp]
\centering
{\renewcommand{\arraystretch}{1.12}
\renewcommand{\tabcolsep}{0.18cm}
\begin{tabular}[c]{c|c|c}
HD-V2 film layer & $\epsilon_{p, \rm max}$ (MeV) & $\epsilon_{p,\rm dep}$ (MeV/proton) \\
\hline
1  & 3.8 & 0.54 \\
2 & 4.7 &  0.50\\
3 & 5.8 & 0.47 \\
4 & 6.8 &  0.45 \\
5 & 7.7 &  0.42 \\
6 & 8.5 &  0.40
\end{tabular}}
\caption{Characterisation of our RCF film pack, using GEANT-4 simulations. Here, $\epsilon_{p, \rm max}$
is the proton energy at which energy deposition in a given film layer is maximised, and $\epsilon_{p,\rm dep}$ is
the deposited energy for such protons.}
\label{table:GEANT4}
\end{table}  
{Such an estimate is justified (even though protons with a range of speeds deposit their energy in a given layer of the RCF stack) by the following argument. In any given film layer, the energy $\epsilon_{p,\mathrm{dep}}$ deposited by protons with incident energy $\epsilon_p$ drops to $<50\%$ of the peak deposited-energy value given in the final column of Table \ref{table:GEANT4} for $|\epsilon_p-\epsilon_{p, \rm max}| \gtrsim 0.25 \, \mathrm{MeV}$. As a result, the full-half-width-maximum (FHWM) $\Delta \epsilon_{p, \rm max}$ is much smaller than $\epsilon_{p, \rm max}$. Protons with incident energy $\epsilon_p < \epsilon_{p, \rm max} - 0.5 \, \mathrm{MeV}$ do not deposit any energy at all, because they do not reach the film layer. On the other hand, protons with incident energy $\epsilon_p \gtrsim \epsilon_{p, \rm max} + 0.25 \, \mathrm{MeV}$ could, in principle, deposit $20$--$50\%$ of the maximum deposited energy. However, in the experiment, the energy distribution of the proton-imaging beam is exponential, with a characteristic temperature of ${\sim}3$ MeV; as a result, the contribution of protons with higher energies is suppressed further, due to their lower number.}

\begin{figure}
  \centering
\includegraphics[width=0.68\linewidth]{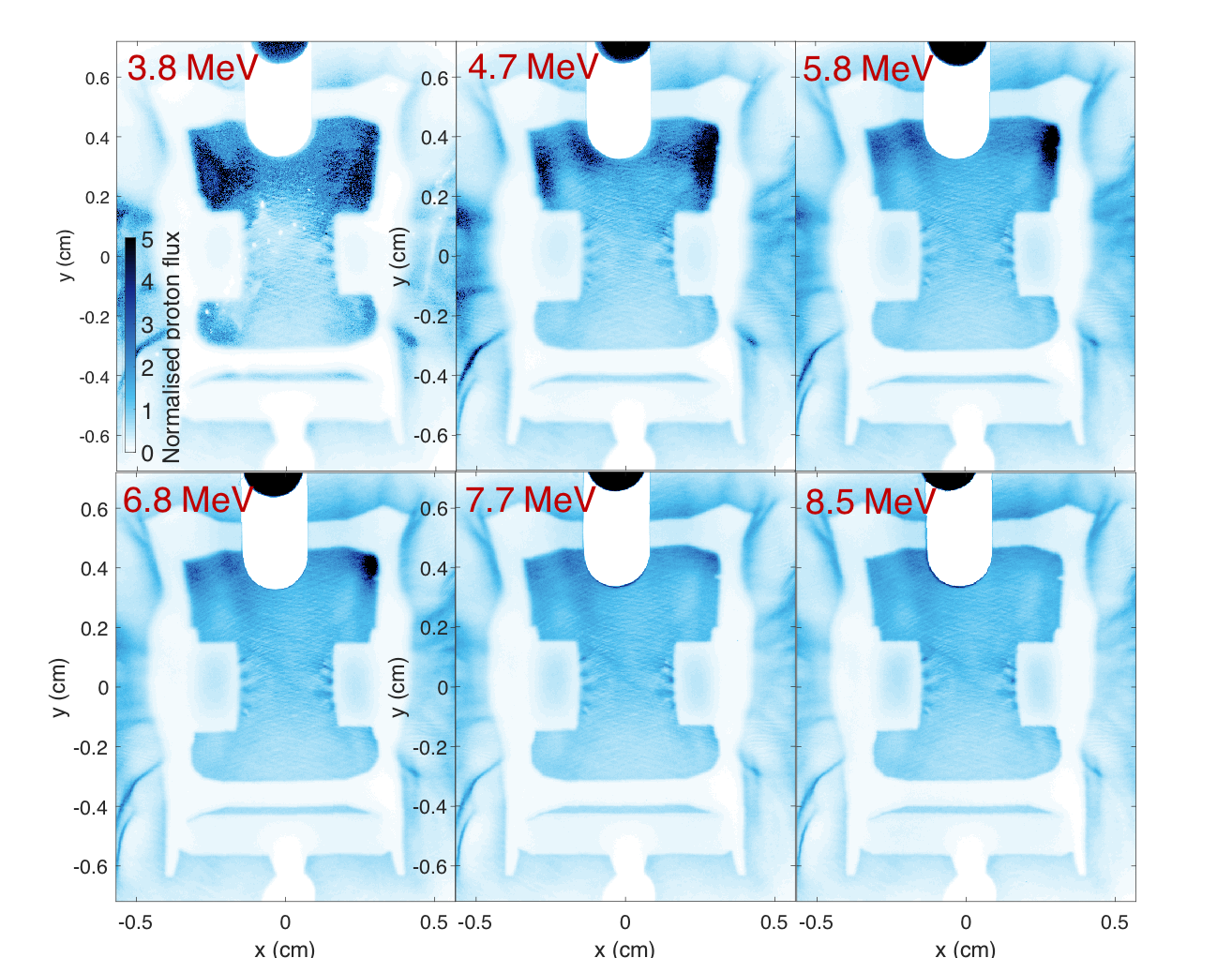}
\caption{\textit{Normalised proton flux images for the first experimental shot}. In this experimental shot, the fastest species of imaging proton
traverses the plasma $\sim 15.7$ ns after the LMJ drive beams are initiated. The energy associated with maximal deposited energy per proton 
is given in the top left-hand corner of each panel. }
\label{fig:protimags_shot1}
\end{figure}

\begin{figure}
  \centering
\includegraphics[width=0.68\linewidth]{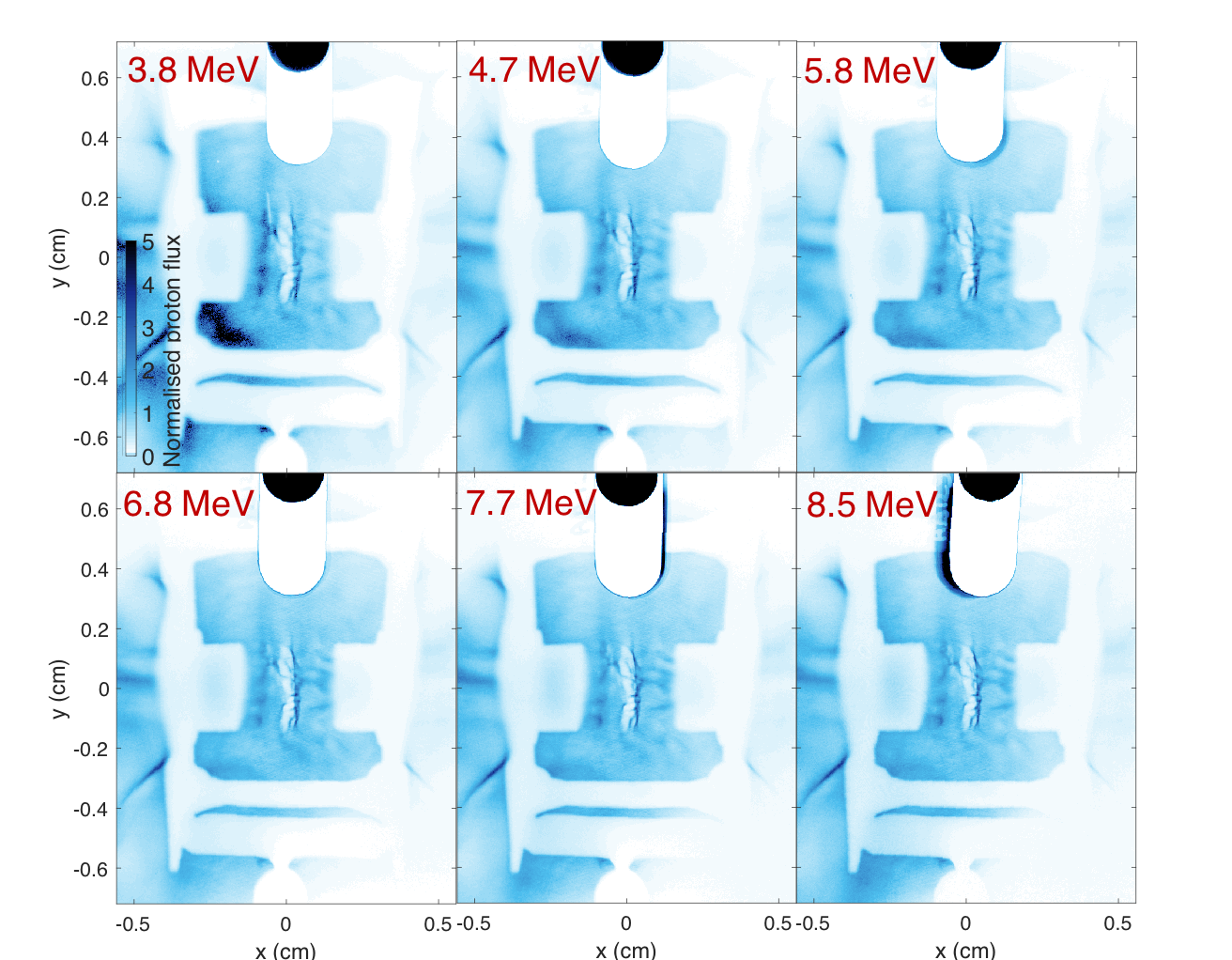}
\caption{\textit{Normalised proton flux images for the second experimental shot}. The same as Supplementary Figure~\ref{fig:protimags_shot1},
except with the fastest species of imaging proton
traversing the plasma $\sim 19.7$ ns after the LMJ drive beams are initiated. }
\label{fig:protimags_shot2}
\end{figure}

\begin{figure}
  \centering
\includegraphics[width=0.68\linewidth]{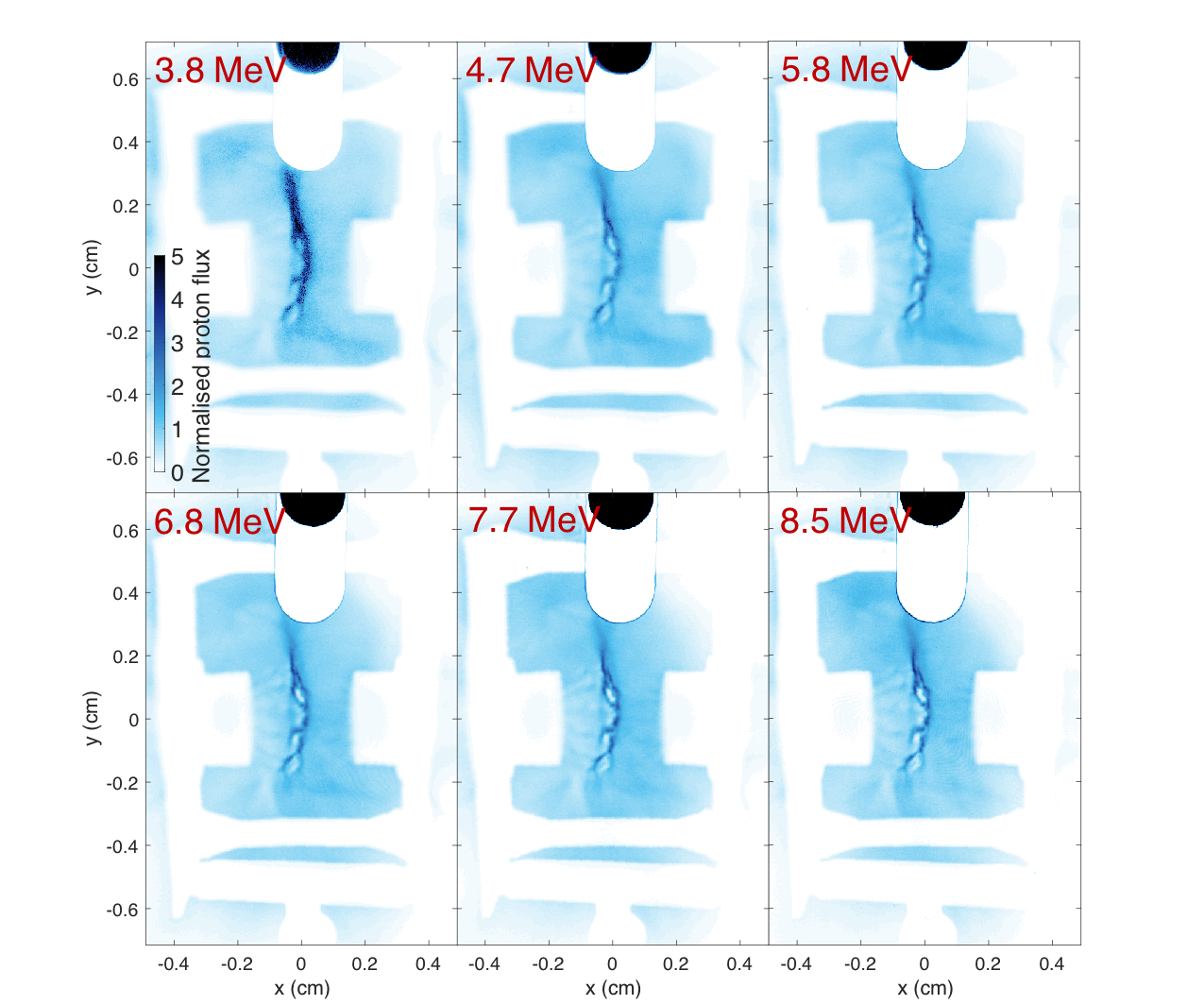}
\caption{\textit{Normalised proton flux images for the third experimental shot}. The same as Supplementary Figure~\ref{fig:protimags_shot1},
except with the fastest species of imaging proton
traversing the plasma $\sim 22.7$ ns after the LMJ drive beams are initiated.}
\label{fig:protimags_shot3}
\end{figure}

The proton images resulting from this (with flux values normalised to the mean value in the region between the grids)
for each of our experimental shots are shown in Figures~\ref{fig:protimags_shot1},~\ref{fig:protimags_shot2} and~\ref{fig:protimags_shot3}. 

\subsection{Recovering path-integrated magnetic fields from proton flux measurements}

In this section, we provide additional details about our analysis of the proton 
images. The reconstruction of the path-integrated magnetic field from a given proton image 
is carried out using the algorithm described in~\citep{AGGS16} (see also the Supplementary Information of~\citep{T18}); 
however, here we clarify a few details of the analysis specific to this experiment. 

First, we explain why we chose to present only the analysis carried out on the proton images arising 
from the sixth RCF film layer (8.5 MeV protons) for all of our experimental shots, and for the 
second RCF film layer (4.7 MeV) for the experimental shot characterizing our experiment 
prior to collision. As stated in the main text, quantitative analysis of a given 
proton image using the approach described in~\citep{AGGS16}  
is only possible if the proton beam does not self-intersect (due to deflections acquired in the plasma) prior 
to reaching the detector. Such self-intersection can be identified by certain 
features present in proton images (see~\citep{AGGS16} for an extended 
discussion). One such feature is the presence of localised proton-flux `structures' 
with two key properties: first, typical values much higher than 
the mean proton flux in an image; second, the broadening of those 
structures in lower-energy proton images of the same fields. Such a feature
can indeed be identified in our proton images for the (two) experimental shots subsequent to the collision 
of the jets in our experiment. This 
suggests that so-called `caustics' are present in all but the highest-energy proton 
image, and thus the quantitative analysis technique proposed in~\citep{AGGS16} cannot be
carried out reliably on the proton images derived from the first five film 
layers. By contrast, the variations in proton flux for the experimental shot prior to the 
jet collision do not have this feature, and so all film layers can, in 
principle, be analyzed. However, the first layer (see Supplementary Figure~\ref{fig:protimags_shot1}, top left) has 
other features suggestive of the film layer being damaged: so a quantitative 
analysis of such an image would likely lead to erroneous results. As the protons detected in second film layer 
arrive with the greatest time difference from the sixth layer, we chose to focus 
our analysis on these two layers for this shot. 

Next, we discuss our approach to characterizing the initial proton beam's flux distribution -- necessary input into 
any magnetic-field-reconstruction algorithm. 
It is known that proton beams arising from high-intensity laser sources can have 
  significant spatial variation prior to any interaction with electromagnetic 
  fields~\citep{N09}. However, such variation is typically on much larger scales than strong 
  variations in proton flux -- so we assume that all variations on scales larger 
  than the known physical scales of interest (i.e., the interaction region's size, $\ell_{n\perp} \approx 0.25 \, \mathrm{cm}$) 
  are variations in the initial proton flux. We determine these initial 
  variations in a two-step procedure: we first calculate a (two dimensional) linear best fit; we then apply a low-pass filter to the difference between the flux distribution and the linear fit 
  (with characteristic filtering scale $\ell = 1.5 \ell_{n\perp}$). The result 
  is illustrated in Supplementary Figure~\ref{fig:initial_variation_analysis}.
\begin{figure}
  \centering
\includegraphics[width=0.5\linewidth]{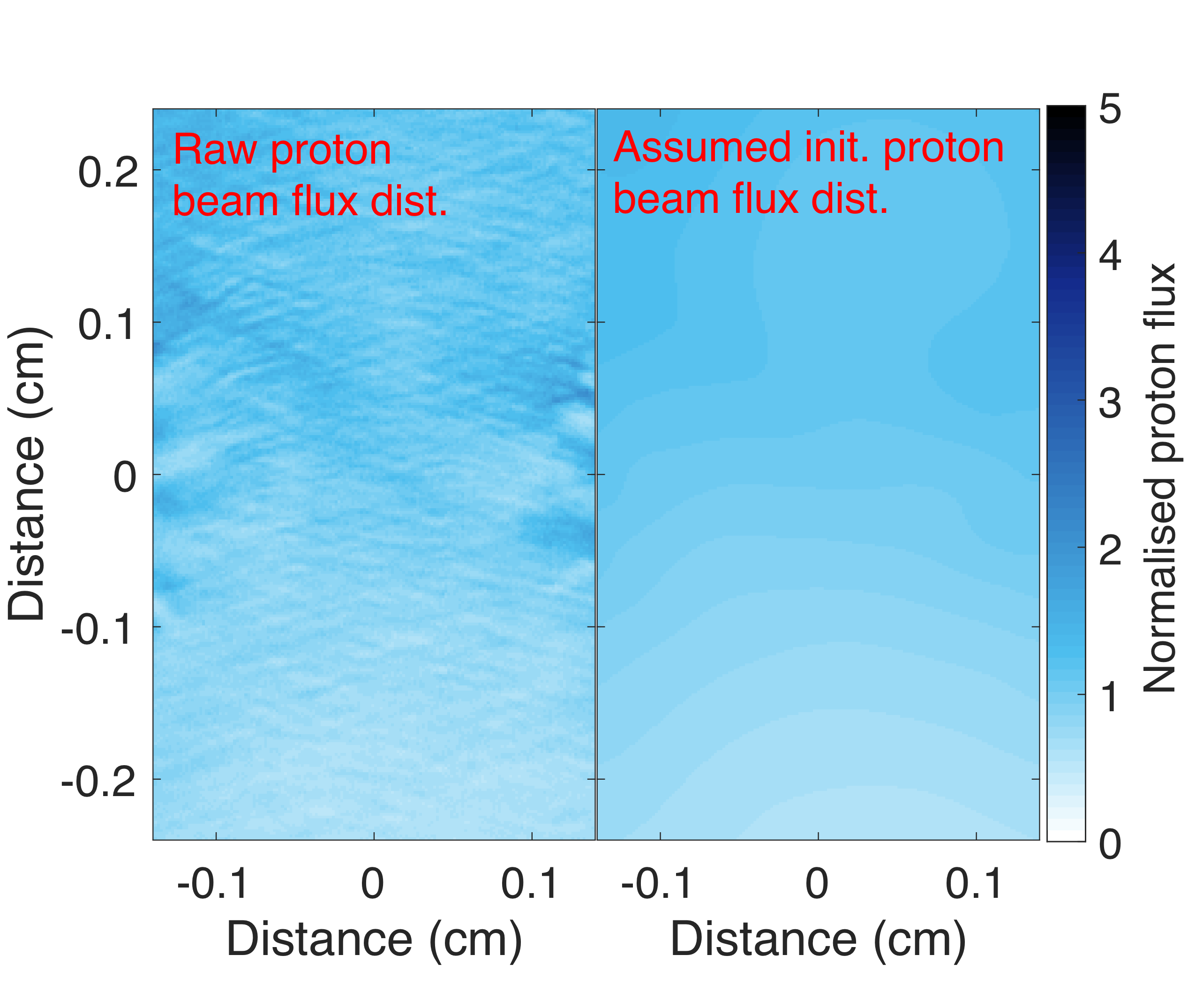}
\caption{\textit{Assumed initial proton-beam flux distribution}. Left: rectangular sample from the 8.5 MeV proton image at $\sim$15.7 ns. The sample is the same as the region depicted in the bottom left of Figure 1 of the main text. 
Right: assumed initial 8.5 MeV proton-beam flux distribution employed for recovering path-integrated magnetic field (which is calculated using the approach described in the accompanying text).}
\label{fig:initial_variation_analysis}
\end{figure}
  The path-integrated field is then reconstructed using the usual technique described in~\citep{AGGS16}; in order to be consistent 
  with our previous assumptions, we apply a Gaussian high-pass filter to the path-integrated 
  fields with the same parameters at the low-pass filter applied to the original 
  proton image. 
  
  Thirdly, we briefly discuss the implications of assuming that the measured proton images are created using a monoenergetic beam (with energy $\epsilon_{p,\mathrm{max}}$) when we apply the field-reconstruction algorithm, despite our statement in the previous section that protons with a distribution of energies (the characteristic dispersion of that energy around the mean, $\Delta \epsilon_{p,\mathrm{max}}$, being $\Delta \epsilon_{p,\mathrm{max}} \approx 0.5$--$1$ MeV) contribute significantly to the deposited energy in each film layer of the RCF film stack. The uncertainty in the imaging protons' energy leads to an uncertainty in the deflection angles of the protons composing a particular proton image; for deflections induced by magnetic fields (where the deflection angle, $\alpha_p$, satisfies $\alpha_{p} \propto \epsilon_p^{-1/2}$), this uncertainty is of order ${\sim}\Delta \epsilon_{p,\mathrm{max}}/2 \epsilon_{p,\mathrm{max}}$. Therefore, the path-integrated magnetic field measurements derived from the second film layer have an ${\sim}15\%$ intrinsic uncertainty due to this effect, while those from the sixth film layer have an ${\sim}5\%$ uncertainty. The latter error is a small fraction of the error (${\sim}20$--$30\%$) of magnetic-field measurements of the interaction-region plasma (which are all derived from the sixth film layer) arising from other sources of uncertainty in the experiment, suggesting that the monoenergetic assumption is a reasonable one for us to make.  
  
  The `total' path-integrated magnetic fields recovered from the proton 
  images obtained after collision of the plasma jets are shown in Supplementary Figure~\ref{fig:total_pathint_field}. 
  \begin{figure}
  \centering
\includegraphics[width=0.5\linewidth]{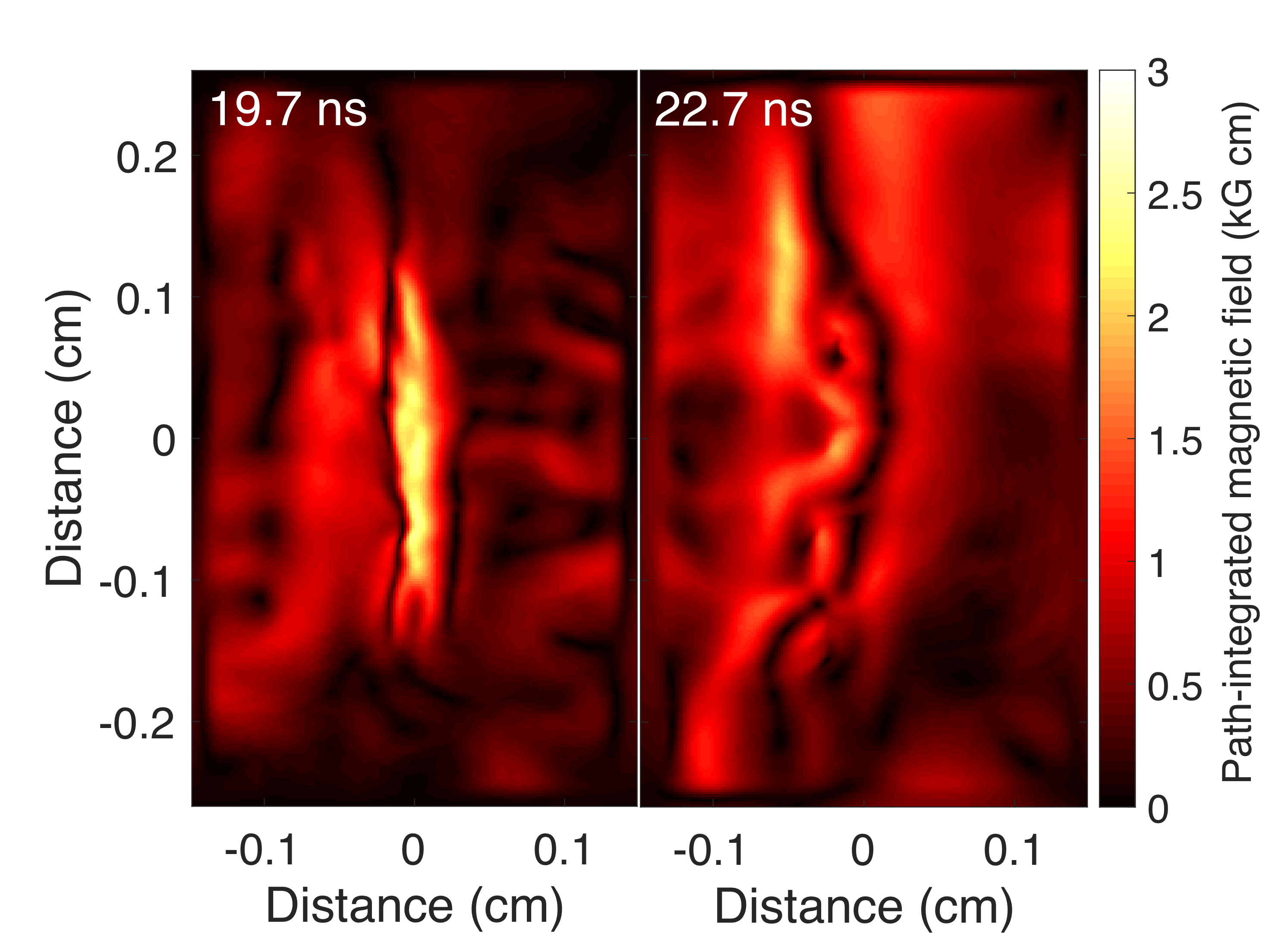}
\caption{\textit{Magnitude of total path-integrated magnetic fields recovered from post-collision proton images}. Left: total path-integrated 
magnetic field recovered from 8.5 MeV proton image taken at $\sim$19.7 ns after the initiation of the LMJ drive beams. The scale parameters 
$\ell_{\|}$ and $\ell_{\perp}$ of the anisotropic Gaussian low- and high-pass filters used to recover `large-scale' and `small-scale'  
components of the path-integrated field (shown in Figure 4 of the main text) are given by $\ell_{\|} = 0.018 \, \mathrm{cm}$, $\ell_{\perp} = 0.15 \, \mathrm{cm}$. 
Right: total path-integrated magnetic field recovered from 8.5 MeV proton image taken at $\sim$22.7 ns. The anisotropic Gaussian filter parameters used in this case 
are $\ell_{\|} = 0.088 \, \mathrm{cm}$, $\ell_{\perp} = 0.2 \, \mathrm{cm}$.}
\label{fig:total_pathint_field}
\end{figure}
  In the main text, 
  we report `large-scale' -- that is, at the scale of the interaction region -- and `small-scale' -- at the characteristic scale of the turbulence, and at smaller scales -- path-integrated 
  fields. In order to recover the large-scale (small-scale) path-integrated magnetic fields from the total path-integrated  
  fields, we apply an additional low-pass (high-pass) filter. The choice of 
  filter is complicated by the fact that the morphology of the 
  interaction region is anisotropic: its parallel size $\ell_{n\|} \approx 400 \, \mu 
  \mathrm{m}$ is much smaller than its perpendicular size ($\ell_{n\perp} \approx 2.5 \, 
  \mathrm{mm}$). We therefore apply an anisotropic Gaussian low-pass filter (scale parameters $\ell_{\|}$ and $\ell_{\perp}$ at particular times given in the caption of Supplementary Figure~\ref{fig:total_pathint_field})
  to obtain the large-scale field, and an anisotropic Gaussian high-pass filter (same parameters) to extract the small-scale field.
  We note that since the parallel size of the interaction region is rather 
  similar to the scale associated with the peak wavenumber of the magnetic-energy 
  spectra, it is possible that some magnetic structures associated with the supersonic turbulence whose associated 
  wavevectors are oriented in the parallel direction will be suppressed by the 
  high-pass filter. However, for an isotropic stochastic magnetic field, the 
  resulting underestimate of the RMS value of the field will only be a small correction (of order $\sim 
 \ell_{\|}/\ell_{\perp} \lesssim 15\%$). 

\begin{figure}
  \centering
\includegraphics[width=0.75\linewidth]{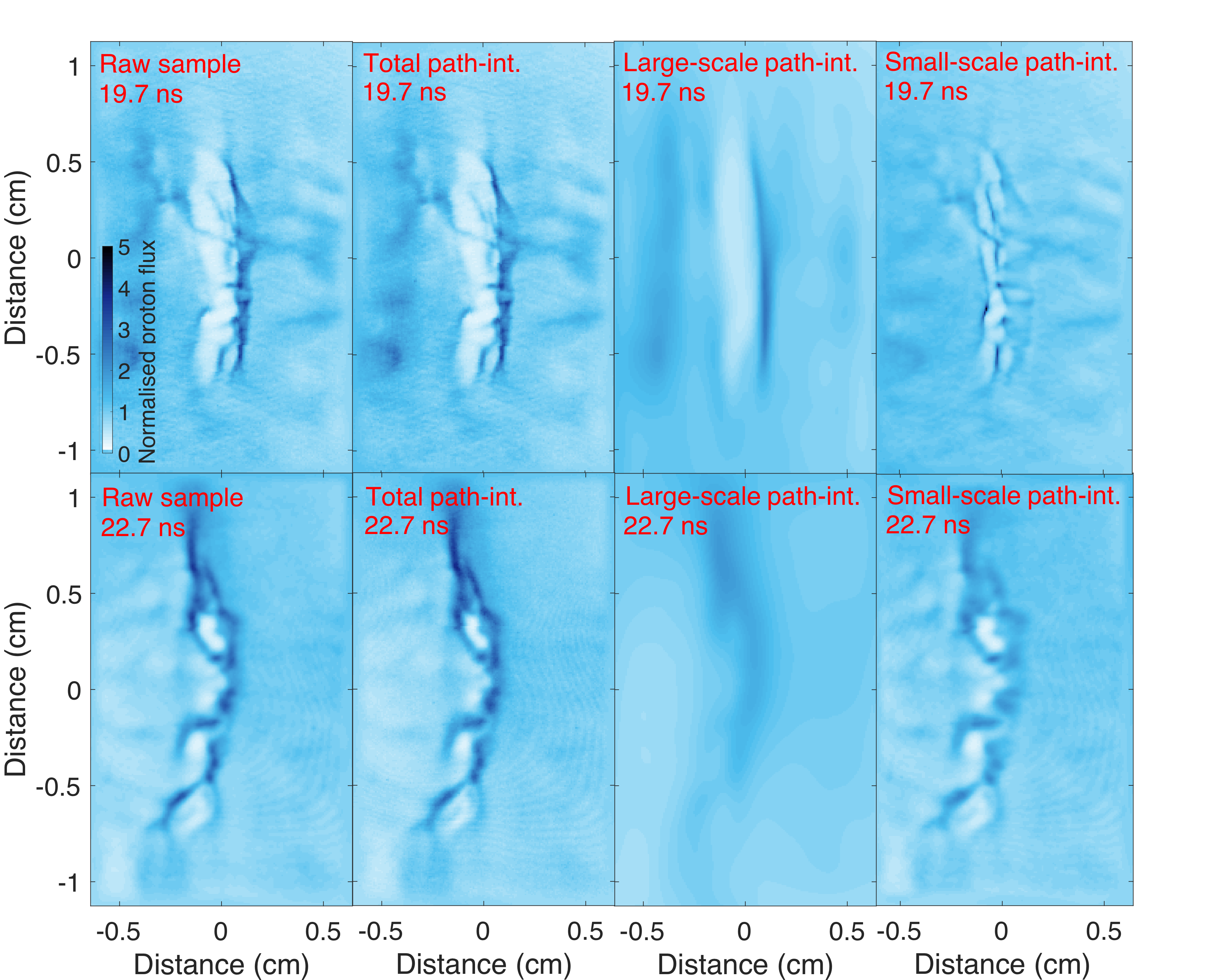}
\caption{\textit{Predicted 8.5 MeV proton images resulting from recovered path-integrated magnetic fields}. Top far-left: sample from 19.7 ns, 8.5 MeV proton image, used for recovering the path-integrated magnetic field.  Top mid-left: 
proton-flux distribution arising from the total path-integrated magnetic field at 19.7 ns (which is presented in Supplementary Figure \ref{fig:total_pathint_field}, left). Top mid-right: proton-flux distribution arising from the large-scale path-integrated field at 19.7 ns (which is presented in Figure 4 of the main text, top left). 
Top far-right: proton flux distribution arising from the small-scale path-integrated field at 19.7 ns (presented in Figure 4 of the main text, top mid-left). Bottom far-left: sample from 22.7 ns, 8.5 MeV proton image. Bottom mid-left:
proton-flux distribution arising from the total path-integrated field at 22.7 ns (presented in Supplementary Figure \ref{fig:total_pathint_field}, right). Bottom mid-right: 
proton-flux distribution arising from the large-scale path-integrated field at 22.7 ns (presented in Figure 4 of the main text, bottom left). 
Bottom far-right: proton-flux distribution arising from the small-scale path-integrated field at 22.7 ns (presented in Figure 4 of the main text, bottom 
mid-left).
}
\label{fig:predicted_prot_images}
\end{figure}

To illustrate how the large-scale and small-scale path-integrated magnetic fields relate back to 
 the proton images from which they were extracted, we show the 
 proton images resulting from the large- and small- scale fields individually in Figure~\ref{fig:predicted_prot_images}. 
 For comparison, we also show the original proton images, and the 
 predicted images associated with the total path-integrated magnetic field. 
It is qualitatively clear that the stochastic, smale-scale flux structures in the original proton 
images are indeed reproduced by imaging the small-scale path integrated field; 
however, the presence of the large-scale field distorts both their shape and 
position. 

\subsection{Electron number density measurements using collisional scattering of imaging protons}

Here, we elaborate on how we obtained the measurements of electron number density of the interaction-region plasma using collisional scattering of 
the proton imaging beam; we also provide a bound on the electron number density of the initial supersonic plasma jets on account of the absence of such scattering.

First, we outline the relationship between the effective proton image resolution 
and the plasma's electron number density. 
A beam of protons (initial speed $V_0$) travelling through a plasma with electron 
number density $n_e$ diffuses in the direction(s) transverse to its initial motion 
due to small-angle Coulomb collisions. The characteristic spread in velocities $\Delta v_{\perp}$  
in the proton beam associated with these collisions evolves according to
\begin{equation}
 \frac{\mathrm{d} \Delta v_{\perp}^2}{\mathrm{d} t} = \left(\nu_{\perp}^{p| \rm Al} + \nu_{\perp}^{p |e}\right) 
 v^2 \, , \label{coll_scat_exact}
\end{equation}
where $v$ is the speed of the proton beam after time $t$ of its interaction with 
the plasma, $\nu_{\perp}^{p| \rm Al}$ is the characteristic perpendicular diffusion rate due to
Coulomb collisions with aluminium ions, and $\nu_{\perp}^{p|e}$ is the characteristic perpendicular diffusion rate due to
Coulomb collisions with electrons~\citep{H94}. 
For the protons used in our experiment, whose velocity is $V_0 \gtrsim 3 \times 10^{9} \, \mathrm{cm/s}$, 
the beam protons' velocities greatly exceed both the plasma's thermal ion and electron 
velocities (assuming a jet-plasma temperature $T \approx 100$ eV); as a 
consequence, $\nu_{\perp}^{p|\rm Al}$ and $\nu_{\perp}^{p|e}$ are given by
the fast test particle rates:
\begin{eqnarray}
\nu_{\perp}^{p|\rm Al} & \approx &  1.8 \times 10^{-7} \epsilon_p^{-3/2} \langle Z \rangle^2 n_i \log{\Lambda_{p|\rm Al}} 
\, , \\
\nu_{\perp}^{p|e} & \approx &  1.8 \times 10^{-7} \epsilon_p^{-3/2} n_e \log{\Lambda_{p|e}} 
\, , \label{coll_scat_rates}
\end{eqnarray}
where $\epsilon_p$ is the proton energy (in eV), $n_i$ the aluminium ion number density, $n_e$ the electron number density, $\langle Z \rangle$ the mean aluminium charge, $\log{\Lambda_{p|\rm Al}}$ the proton-ion Coulomb logarithm, and $\log{\Lambda_{p|e}}$ 
the proton-electron Coulomb logarithm. 
If Coulomb collisions are 
sufficiently weak for the protons' velocities to be only slightly perturbed 
before they traverse the whole plasma (assumed path length $\ell_{n\perp}$), it 
follows from (\ref{coll_scat_exact}) that 
\begin{equation}
\Delta v_{\perp} \approx V_0  \sqrt{\left(\nu_{\perp}^{p| \rm Al} + \nu_{\perp}^{p|e}\right) t_{\rm cross}} \approx \sqrt{\left(\nu_{\perp}^{p|\rm Al} + \nu_{\perp}^{p|e}\right) \ell_{n\perp} V_0}  
\, ,
\end{equation}
where $ t_{\rm cross} \approx \ell_{n\perp}/V_0$ is the time taken by the proton beam to traverse 
the plasma. 
The characteristic scattering angle $\Delta \theta_{\rm coll}$ associated with 
collisional interactions is then given by
\begin{equation}
\Delta \theta_{\rm coll} \approx \frac{\Delta v_{\perp}}{V_0} \approx \sqrt{\frac{\left(\nu_{\perp}^{p|\rm Al} + \nu_{\perp}^{p|e}\right) \ell_{n\perp}}{V_0}} 
\, . \label{coll_scat_angle}
\end{equation}
For any proton scattering process with characteristic scattering angle $\Delta \theta \ll 1$, it can be shown using kinetic 
theory~\citep{AGGS16}
that the proton-flux distribution measured on the detector in a proton-imaging diagnostic set-up 
is the convolution of the proton-flux distribution in the absence of that 
scattering process with a Gaussian point-spread function, whose full-half-width maximum (FHWM) $a_{\rm res}$ is given by $a_{\rm res} \approx 2.2 r_s \Delta \theta$, where $r_s$ is the distance 
from the plasma to the detector. Assuming the scattering process is collisional 
scattering, (\ref{coll_scat_angle}) gives
\begin{equation}
a_{\rm res} \approx 2.2 r_s  \sqrt{\frac{\left(\nu_{\perp}^{p|\rm Al} + \nu_{\perp}^{p|e}\right) \ell_{n\perp}}{V_0}}  
\, .
\end{equation}
Now substituting (\ref{coll_scat_rates}), we obtain
\begin{equation}
a_{\rm res} \approx 0.045 \left[\frac{r_s}{10 \, \mathrm{cm}}\right] \left[\frac{\epsilon_p}{4.7 \, \mathrm{MeV}}\right]^{-1} \left[\frac{\ell_{n\perp}}{0.25 \, \mathrm{cm}}\right]^{1/2} \left[\frac{n_e}{4 \times 10^{19} \, \mathrm{cm}^{-3}}\right]^{1/2} \left[\frac{\langle Z \rangle}{11} \frac{\log{\Lambda_{p|\rm Al}}}{6} + \frac{\log{\Lambda_{p|e}}}{6}\right]^{1/2} 
\, \mathrm{cm} , 
\end{equation}
where $\langle Z \rangle$ and the Coulomb logarithms are calculated using the 
characteristic parameters of the beam and plasma. Taking account of the  
proton-imaging magnification factor $\mathcal{M} = 13/3$, this means that wavenumber cutoff $k_{\rm res}$ 
due to collisional broadening is given by 
\begin{equation}
k_{\rm res} \approx 4.4 \mathcal{M}/a_{\rm res} \approx 400 \, 
\mathrm{cm}^{-1} \, .
\end{equation}

With the relationship between the proton-image resolution and electron number 
density determined, we now explain how we 
construct a predictive model of the high-wavenumber tail of the spectrum of the sharp, large-amplitude proton-flux inhomogeneities (known as caustics) in the 4.7 MeV proton images at a given electron number density. We first calculate the one-dimensional spectrum $E_{\delta \Psi}(k)$ of the relative flux distribution -- 
that is, the spectrum of $\delta \Psi \equiv (\Psi-\Psi_0)/\Psi_0$, where $\Psi$
is the actual proton-flux distribution, and $\Psi_0$ is the mean proton flux -- 
for a particular region of the 4.7 MeV proton images (the regions used are shown in Figure 3 of the main 
text). We then assume that the measured spectrum can be expressed in the 
following manner:
\begin{equation}
E_{\delta \Psi}^{4.7 \, \mathrm{MeV}}(k) = E_{\hat{\delta \Psi}}^{4.7 \, \mathrm{MeV}}(k) \exp{\left[-\frac{k^2}{k_{\rm res}^2(\epsilon_p = 4.7 \, \mathrm{MeV})}\right]} 
+ E_{\rm noise}(\epsilon_p = 4.7 \, \mathrm{MeV})
 \, , \label{rel_flux_spec_ansatz_4MeV}
\end{equation}
where $E_{{\delta \Psi}}^{4.7 \, \mathrm{MeV}}(k)$ is the measured spectrum from the 4.7 MeV proton image, $E_{\hat{\delta \Psi}}^{4.7 \, \mathrm{MeV}}(k)$ is the `true' spectrum of the large-amplitude flux inhomogeneities in the absence of both 
collisions and image noise in the 4.7 MeV proton image, $k_{\rm res}(\epsilon_p = 4.7 \, \mathrm{MeV})$ is the collisional wavenumber cutoff for 4.7 MeV protons, and $E_{\rm noise}(\epsilon_p = 4.7 \, \mathrm{MeV})$ is the noise level 
for the 4.7 MeV proton image. Next, we use the fact that the high-wavenumber tail of the true spectrum of caustics (with characteristic separation scale $\ell_{\Psi} \gg k^{-1}$) is given by $E_{\hat{\delta \Psi}}^{4.7 \, \mathrm{MeV}}(k) \approx E_{\hat{\delta \Psi},0} (k \ell_{\Psi})^{-1}$, where $E_{\hat{\delta \Psi},0}$ is a normalisation constant chosen so that the the low-wavenumber part of the measured and true spectra coincide~\citep{AGGS16}. We can then use (\ref{rel_flux_spec_ansatz_4MeV}) to give an 
expression for $E_{\delta \Psi}^{4.7 \, \mathrm{MeV}}(k)$ in terms of the two constants $E_{\hat{\delta \Psi},0}$ and $E_{\rm noise}(\epsilon_p = 4.7 \, \mathrm{MeV})$, which can be determined directly from the data, and $k_{\rm res}(\epsilon_p = 4.7 \, \mathrm{MeV})$, which is a function of $n_e$:
\begin{equation}
E_{\delta \Psi}^{4.7 \, \mathrm{MeV}}(k) \approx \frac{ E_{\hat{\delta \Psi},0}}{k \ell_{\Psi}} \exp{\left[-\frac{k^2}{k_{\rm res}^2(\epsilon_p = 4.7 \, \mathrm{MeV})}\right]} 
+ E_{\rm noise}(\epsilon_p = 4.7 \, \mathrm{MeV}) \, . 
 \label{rel_flux_spec_ansatz_4MeV_B}
\end{equation}
This is the model used in Figure 3 of the main text to predict the 4.7 MeV relative flux spectrum for a particular electron number density.

\begin{figure}
  \centering
\includegraphics[width=0.9\linewidth]{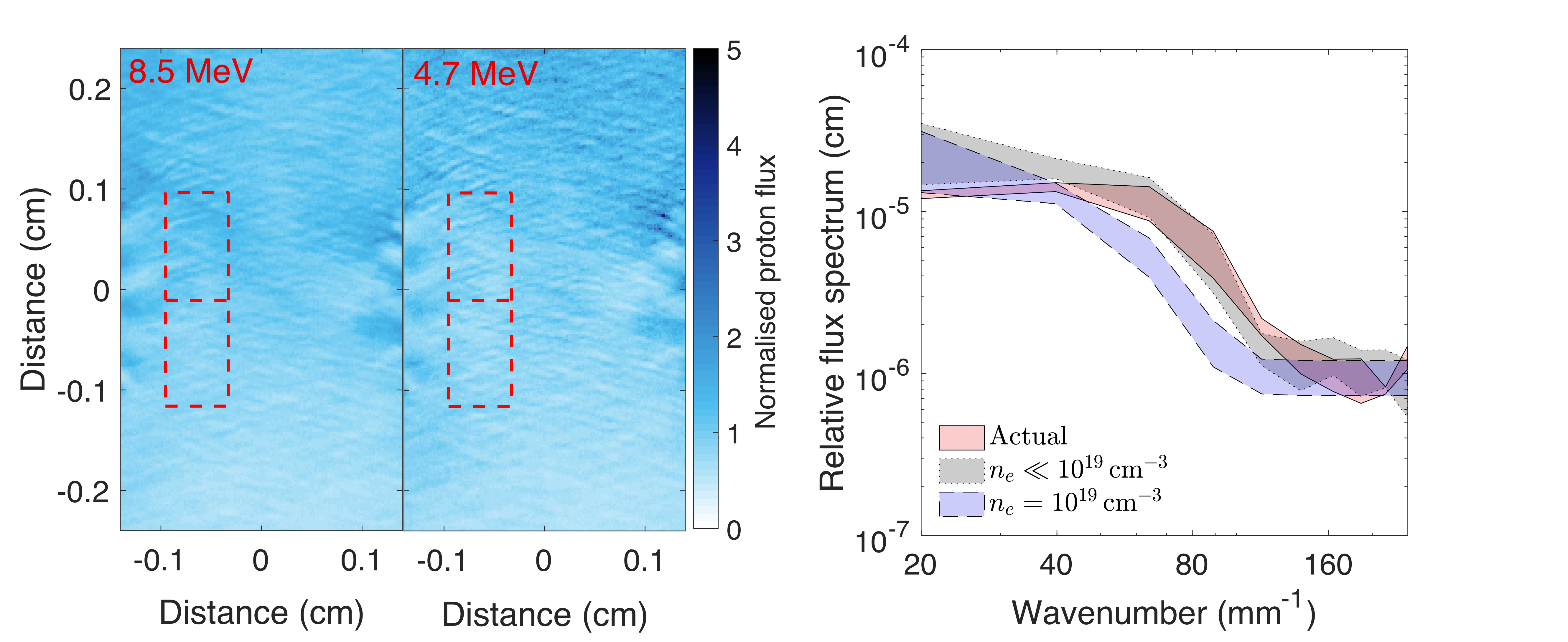}
\caption{\textit{Upper bound on the electron number-density}. Left: samples from 8.5 MeV (far left) and 4.7 MeV (middle left) proton images. Right: 
spectrum of relative 4.7 MeV proton flux (red), as well as the predicted spectra determined using the 8.5 MeV relative-proton-flux spectrum combined with collisional broadening assuming $n_e = 10^{19} \, \mathrm{cm}^{-3}$ (blue), and with negligible broadening (black). The mean and error for each spectrum are 
calculated by combining the individual results from the regions demarcated by the dashed red lines.} 
\label{fig:denstemp}
\end{figure}

In addition to our measurements of the electron number density $n_e$ of the interaction-region plasma, we can also place an upper 
bound on $n_e$ in each plasma jet prior to collision, based on the absence of collisional scattering of the proton imaging beam. The effective resolution of the 4.7 MeV 
pre-collision proton images (Figure \ref{fig:denstemp}, middle) was the same as for the 8.5 MeV images
 (Figure \ref{fig:denstemp}, left). 
 The characteristic magnitude of the small-scale flux inhomogeneities evident in both 
 images -- likely associated with electromagnetic collisionless microinstabilities 
 arising at early times in the 
 experiment~\citep{H15}, before plasma densities rise sufficiently for collisional damping 
 to suppress them
 -- is small, and thus, in the absence of collisions of the proton beam,
 the Fourier spectrum $E_{\hat{\delta \Psi}}(k)$ of the flux inhomogeneities in both images is the same, up to a scaling factor 
 equal to the ratio of the beam-proton energies (viz., $E_{\hat{\delta \Psi}}(k) \propto \epsilon_p^{-1}$)~\citep{AGGS16}. This is indeed the case for our data (see Figure \ref{fig:denstemp}, right). Writing down an expression analogous to (\ref{rel_flux_spec_ansatz_4MeV}) for the 8.5 MeV protons,
 \begin{equation}
 E_{\delta \Psi}^{8.5 \, \mathrm{MeV}}(k)  = E_{\hat{\delta \Psi}}^{8.5 \, \mathrm{MeV}}(k) \exp{\left[-\frac{k^2}{k_{\rm res}^2(\epsilon_p = 8.5 \, \mathrm{MeV})}\right]} 
+ E_{\rm noise}(\epsilon_p = 8.5 \, \mathrm{MeV}) \, ,
 \end{equation}
 we can then construct a model for the measured spectrum of 
 small-scale flux inhomogeneities in the 4.7 MeV proton images, 
given a particular electron number density and the equivalent spectrum in the 
8.5 MeV proton images:
\begin{eqnarray}
E_{\delta \Psi}^{4.7 \, \mathrm{MeV}}(k) & \approx & 1.8 \left[ E_{\delta \Psi}^{8.5 \, \mathrm{MeV}}(k) - E_{\rm noise}(\epsilon_p = 8.5 \, \mathrm{MeV})\right] \nonumber \\
&& \times \exp{\left[\frac{k^2}{k_{\rm res}^2(\epsilon_p = 8.5 \, \mathrm{MeV})}-\frac{k^2}{k_{\rm res}^2(\epsilon_p = 4.7 \, \mathrm{MeV})}\right]} 
+ E_{\rm noise}(\epsilon_p = 4.7 \, \mathrm{MeV}) \, . 
 \label{rel_flux_spec_model_smallamp}
\end{eqnarray}
For electron number densities $n_e \geq 10^{19} \, \mathrm{cm}^{-3}$, Figure \ref{fig:denstemp}, right, shows that the high-wavenumber tail of the 4.7 MeV proton-flux spectrum would be suppressed; we therefore conclude from the absence
 of any suppression that $n_e \ll 10^{19} \, \mathrm{cm}^{-3}$.  This is 
 consistent with the estimates of the electron number density obtained in the FLASH 
 simulation, in which $n_e \approx 10^{18} \, \mathrm{cm}^{-3}$ at the front of each jet~\citep{KT20}.

\section{Plasma characterization}

\subsection{Calculation of interaction-region plasma parameters}
Table \ref{table:TableTheoParam} presents a summary (both formulae and values) of key plasma parameters for 
our experiment, including all those referenced in the main text. The formulae 
used are derived from~\citep{H94,B65,C14}; the opacities are calculated using data tables given by~\citep{ALANL}.  
\begin{table}[htbp]
\centering
{\renewcommand{\arraystretch}{1.12}
\renewcommand{\tabcolsep}{0.18cm}
\begin{tabular}[c]{c|c|c}
Quantity & Formula & Value \\
\hline
Aluminium mass ($M$) & & 27 \\
Mean aluminium charge ($\langle Z \rangle$) & & $\sim$11 \\
Temperature ($T$) & & $100$ eV \\
Electron number density ($n_{e}$) & & $7 \times 10^{19}$ cm$^{-3}$ \\
Aluminium number density ($n_{i}$) & & $6 \times 10^{18}$ cm$^{-3}$ \\
Turbulent velocity ($u_{\rm turb}$) & & 2$\times 10^{7}$ cm $\mathrm{s}^{-1}$  \\
Outer scale ($L$) & & 0.04 cm \\
Magnetic field ($B$) & & 10 kG \\
Adiabatic index ($\gamma_{I}$) & & 5/3 \\
Coulomb logarithm $(\log{\Lambda})$ & $23.5 - \log{n_{e}^{1/2} T^{-5/4}}-\sqrt{10^{-5}+{\left(\log{T}-2\right)^2}/{16}}$ & $\sim 6$ \\
Mass density $(\rho)$ & $1.7 \times 10^{-24} M n_{e}/\langle Z \rangle $ & $2.9 \times 10^{-4}$ g cm$^{-3}$ \\
Debye Length ($\lambda_{\mathrm{D}}$) & $ 7.4 \times 10^2 \,{T^{1/2}}\left(1+ \langle Z \rangle \right)^{-1/2} {n_{{e}}^{-1/2}} $ & \textcolor{black}{$2.5 \times 10^{-7}$ cm} \\
Sound speed ($c_{\mathrm{s}}$) & $ 9.8 \times 10^5 \, {\left[(\langle Z \rangle +1) \gamma_I T \right]^{1/2}}{M^{-1/2}}$  & \textcolor{black}{$8 \times 10^{6}$ cm s$^{-1}$} \\
Mach number & $u_{\rm turb}/c_{\mathrm{s}}$  & $2.5$ \\
Plasma $\beta$ & $4.0 \times 10^{-11} \, {\left( 1+ \langle Z \rangle^{-1} \right) n_e T}/{B^2} $ & $3 \times 10^3$\\
Ion-ion mean free path ($\lambda_{ii}$) & $2.9 \times 10^{13} \, {T^2}/{\langle Z \rangle^{3} n_{e} \log{\Lambda}}$ & \textcolor{black}{$5 \times 10^{-7}$ cm}\\
Electron-ion mean free path ($\lambda_{e}$) & $2.1 \times 10^{13} \, {T^2}/{\langle Z \rangle n_{{e}} \log{\Lambda}}$ & \textcolor{black}{$5 \times 10^{-5}$ cm}\\
Electron-ion equilibration time ($\tau_{ie}^{\epsilon}$) & $3.1 \times 10^8  \, {M T^{3/2}}/{\langle Z \rangle n_{e}  \log{\Lambda} }$ & \textcolor{black}{$4 \times 10^{-9}$ s} \\
Electron Larmor radius ($\rho_{e}$) &\textcolor{black}{ $2.4  \, {T^{1/2}}/{B}$ }& \textcolor{black}{$2.4 \times 10^{-3}$ cm}\\
Ion Larmor radius ($\rho_{i}$) & \textcolor{black}{ $1.0  \times 10^2 \, {M^{1/2} T^{1/2}}/{\langle Z \rangle B}$} & \textcolor{black}{$5 \times 10^{-2}$ cm} \\
Thermal diffusivity ($\chi$) & $4 \times 10^{21} \, {T^{5/2}}/{\langle Z \rangle n_{{e}} \log{\Lambda}}$& $9.0 \times 10^{4}$ cm$^2$ s$^{-1}$\\
Turbulent Peclet number ($\mathrm{Pe}$) & $u_{\rm turb} L/\chi$ & 9 \\
Kinematic viscosity ($\nu$) & $1.9 \times 10^{19} \, {T^{5/2}}/{M^{1/2} \langle Z \rangle^{3} n_{{e}} \log{\Lambda}}$  & \textcolor{black}{$0.7$ cm$^2$ s$^{-1}$} \\
Turbulent Reynolds number ($\mathrm{Re}$) & $u_{\rm turb} L/\nu $ & $10^6$ \\
Resistivity ($\eta$) & $2.8 \times 10^5 \,  {\langle Z \rangle} \log{\Lambda}/{T^{3/2}}$  & \textcolor{black}{$1.8 \times 10^{4}$ cm$^2$ s$^{-1}$} \\
Magnetic Reynolds number ($\mathrm{Rm}$) & $u_{\rm turb} L/\eta$ & \textcolor{black}{$45$} \\
Magnetic Prandtl number ($\mathrm{Pm}$) & $\mathrm{Rm}/\mathrm{Re}$ & $ 4 \times 10^{-5}$ \\
Planck Opacity ($\kappa_{P}$) & & 30 cm$^2$ g$^{-1}$ \\
Photon mean free path ($\lambda_{P}$) & $1/\rho \kappa_{P}$ & 120 cm\\
Cooling function ($L_{\Lambda}$) & 1$.03 \times 10^{12} \, \rho \kappa_{P} T^4$ & 
$9 \times 10^{17}$ erg cm$^{-3}$ s$^{-1}$\\
Radiative cooling time ($\tau_{\rm rad}$) & $1.4 \times 10^{12} \rho (\langle Z \rangle + 1) T/M L_{\Lambda}$ & 
$2 \times 10^{-8}$ s
\end{tabular}}
\caption{Characteristic parameters in the region of supersonic plasma turbulence.
The units system used for all physical quantities in the above formulas is Gaussian CGS, except for the temperature, which expressed in eV. }
\label{table:TableTheoParam}
\end{table} 

For our reported parameters, the plasma may appear to be well described as being optically thin ($\lambda_{P}$ is much larger than the largest spatial dimension of the 
interaction-region plasma, $\ell_{n\perp} \approx 0.25 \, \mathrm{cm}$), and 
the radiative cooling time is much longer than the turnover time of the plasma 
turbulence. However, we caution that the Planck opacity for aluminium 
increases by around three orders of magnitude over the temperature interval $T = 20$ eV to $T = 100$ eV. As a 
consequence of this, the radiative cooling time for a plasma with the same mass density as
given in Table \ref{table:TableTheoParam}, but with $T = 50$ eV, is considerably 
smaller than that at $T = 100$ eV: $\tau_{\rm rad}(T = 50 \, \mathrm{eV}) \approx 1.7 \, 
\mathrm{ns}$ (see Supplementary Figure~\ref{fig:rad_properties}, left). 
  \begin{figure}
  \centering
\includegraphics[width=0.8\linewidth]{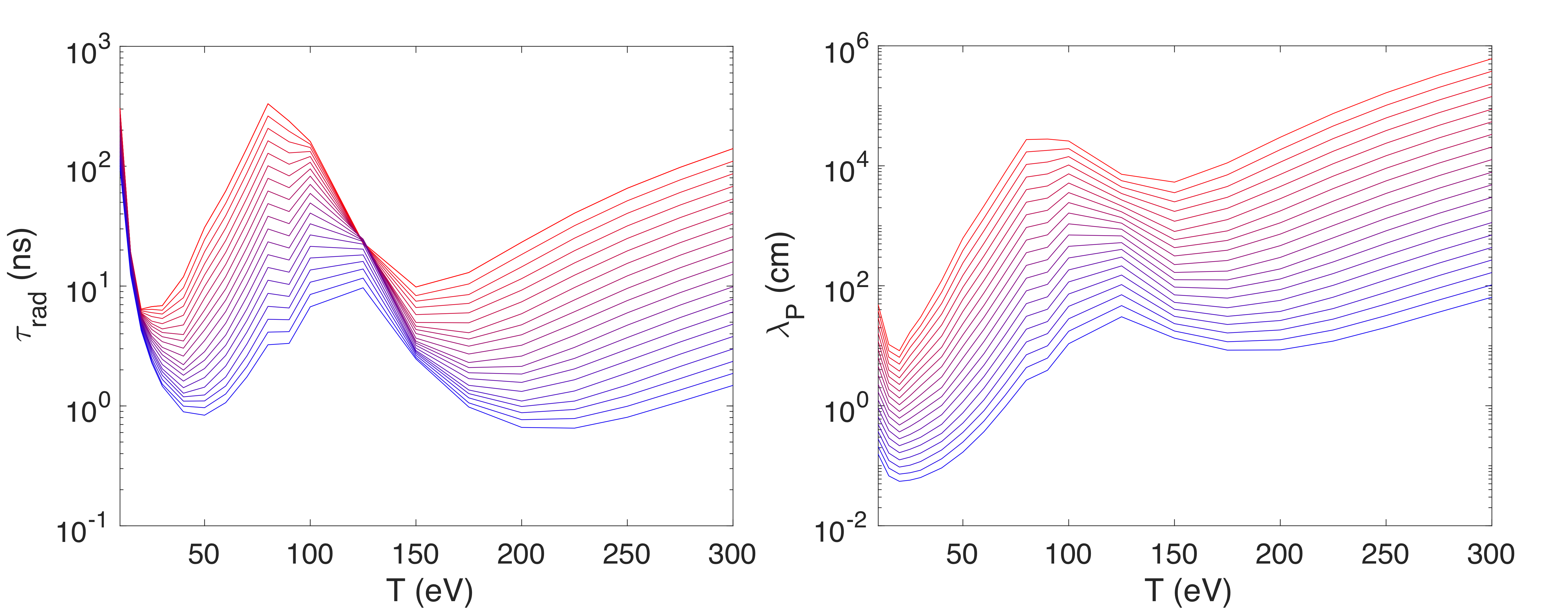}
\caption{\textit{Radiative properties of aluminium plasma relevant to our experiment}. Left: radiative cooling time 
(calculated using the formula given in Supplementary Table \ref{table:TableTheoParam}), plotted as a function of temperature. The 20 different lines plotted 
correspond to this curve at different mass densities, with $\rho_{\rm min} = 10^{-5} \, \mathrm{g} \, \mathrm{cm}^{-5}$, and $\rho_{\rm max} = 10^{-3} \, \mathrm{g} \, \mathrm{cm}^{-3}$, and equal logarithmic spacing.
Right: same as left, but showing photon mean free path.}
\label{fig:rad_properties}
\end{figure}
As well as this, for $T \gtrsim 100 \, \mathrm{eV}$, the Planck opacity increases as the temperature does (until $T \approx 175 \, \mathrm{eV}$); the cooling function increases by nearly two orders of magnitude, giving $\tau_{\rm rad}(T = 175 \, \mathrm{eV}) \approx 1.4 \, 
\mathrm{ns}$. Because the plasma turbulence is both supersonic 
and has a large Peclet number, it is to be expected that there be both significant 
density and temperature variations in the interaction-region plasma; as a 
consequence, we cannot rule out the possibility that radiative cooling plays a 
significant role in the turbulent plasma's dynamics. We do, however, observe 
that the characterisation of the plasma as being optically thin remains appropriate, 
even in the presence of significant temperature variations. At $T = 50$ eV, the 
photon mean free path $\lambda_{P}$ is given by $\lambda_{P} \approx 0.5 \, 
\mathrm{cm}$ (see Figure~\ref{fig:rad_properties}, right), 
which remains much larger than the scale of temperature variations ($L \approx 400 \, \mu \mathrm{m}$). 

\subsection{Plasma parameters of supersonic plasma jets prior to collision}

As discussed in the main text, we do not have a measurement of the temperature $T_{\rm jet}$
of either jet prior to collision. 
However, we can estimate $T_{\rm jet}$ in the following manner, using a combination of physical arguments and bespoke FLASH simulations of the LMJ experiment. The latter, {which assumed a more efficient laser-foil coupling efficiency than was likely realised in practice, and thus obtained greater jet velocities by a factor of ${\sim}2$, } 
suggest that the initial jet Mach number $\mathrm{Ma}_{\rm jet}$ of the jet in the experiment satisfies the following bound: $\mathrm{Ma}_{\rm jet} \lesssim 10$. This being the case, we use the known jet velocity ($u_{\rm jet} \approx 290 \, \mathrm{km/s}$) to estimate
\begin{equation}
  T_{\rm jet} \gtrsim 20 \left[\frac{\gamma_I}{5/3}\right]^{-1} \left[\frac{\langle Z_{\rm jet} \rangle +1}{6}\right]^{-1}  \left[\frac{M}{27}\right]^{1/2} \left[\frac{u_{\rm jet}(\mathrm{km/s})}{290 \, 
  \mathrm{km/s}}\right]^2 \, \mathrm{eV} \, , \label{T_jet_lowerbound}
\end{equation}
where we have used a Saha ionization model to estimate $\langle Z_{\rm jet} \rangle$ at $T = 20$ 
eV. If we combine this bound with the physical requirement that $T_{\rm jet} \ll T = 100$ eV (in other
words, assume that the plasma jets experience significant heating when they collide), 
we conclude that $20 \, \mathrm{eV} \lesssim T_{\rm jet} \lesssim 50 \, 
\mathrm{eV}$. Once these bounds are established, we observe that for the 
characteristic jet densities ($\rho_{\rm jet} \sim 10^{-5} - 10^{-4} \, \mathrm{g/cm}^{3}$), the radiative cooling time $\tau_{\rm rad,jet}$ is approximately $\tau_{\rm rad,jet} \sim 3 - 10$ ns. 
This is comparable to the time taken for the jet to travel the distance $\ell_{\rm grid}  \approx 0.2 \, \mathrm{cm}$ from the 
grids in the centre of the target ($t_{\rm grid} \approx \ell_{\rm grid}/u_{\rm jet} \approx 7 \, 
\mathrm{ns}$). Thus, we conclude that each jet experiences significant cooling as it travels towards its 
counterpart, and so a reasonable estimate for $T_{\rm jet}$ is the lower bound (\ref{T_jet_lowerbound}): $T_{\rm jet} \sim 20 \, 
\mathrm{eV}$. 

Using the formula for $\eta$ given in Table \ref{table:TableTheoParam}, we can now estimate 
the magnetic Reynolds number of either jet: $\mathrm{Rm}_{\rm jet} \approx u_{\rm jet} L/\eta_{\rm jet} \approx 
20$. This justifies the assumption made in the main text that the magnetic field 
is effectively frozen into the flow of each jet over the timescale $\Delta t_p \approx 300 \, \mathrm{ps}$ 
separating the two proton species used for imaging the jets. 

\subsection{The role of the Biermann battery in our experiment}

The Biermann battery is known to generate significant magnetic fields in 
colliding laser-plasma jet experiments; here, we elaborate on their importance 
in our experiment. 
In the main text, we claim that the subsequent amplification of a stochastic component 
of the magnetic field observed in our experiment
can be attributed to the action of supersonic motions, rather than to the
action of the Biermann battery alone. To justify this claim, we consider the induction 
equation governing the evolution of the magnetic field, 
\begin{equation}
  \frac{\partial \boldsymbol{B}}{\partial t} = \nabla \times \left(\boldsymbol{u} \times \boldsymbol{B}\right) 
  - \frac{ c k_{B}}{e n_e} \nabla n_e \times \nabla T_e + \eta \nabla^2 
  \boldsymbol{B} \, ,
\end{equation}
where $k_{B}$ is Boltzmann's constant, $e$ is the elementary charge, and $c$ is the speed 
of light, and compare the respective sizes of the inductive term and the 
Biermann term just after collision:
\begin{equation}
  \frac{|\boldsymbol{u} \times \boldsymbol{B}|}{|c k_B\nabla n_e \times \nabla T_e/e n_e|} 
  \sim  \left[\frac{\delta n_e}{n_e}\right]^{-1} \left[\frac{B_0(\mathrm{kG})}{10 \, \mathrm{kG}}\right] \left[\frac{T_e(\mathrm{eV})}{100 \, \mathrm{eV}}\right]^{-1} \left[\frac{L(\mathrm{cm})}{0.04 \, \mathrm{cm}}\right] \left[\frac{{u}_{\rm turb}(\mathrm{cm} \; \mathrm{s}^{-1})}{1.9 \times 10^{7} \, \mathrm{cm} \; \mathrm{s}^{-1}}\right] 
  \, .
\end{equation}
We conclude that the terms are comparable, provided that both the 
variations in density and in temperature are comparable to their mean values.

\section{Relationship of results to previous experiments}

In the main text, we reference several previous laser-plasma experiments that reported turbulent amplification of magnetic fields~\citep{M15,T18,B20,M20}, and one which reported creating boundary-free supersonic turbulence without observing amplification~\citep{W19}; here, we outline in more detail the relationship between these experiments and the experiment reported in this paper. Table \ref{table:ExpRef} provides a summary of the key physical parameters which were attained in the experiments at times coincident with the observation of magnetic-field amplification.
\begin{table}[htbp]
\centering
{\renewcommand{\arraystretch}{1.12}
\renewcommand{\tabcolsep}{0.18cm}
\begin{tabular}[c]{c|c|c|c|c|c|c|c}
Experiment & Identifier & $L$(cm)& $\mathrm{Rm}$ & $\mathrm{Pm}$ & $\mathrm{Ma}_{\rm turb}$ & $\delta B/B_0$ & $E_{\rm mag}/E_{\rm kin}$  \\
\hline
This paper & LMJ & 0.04 & 45 & ${\sim}4 \times 10^{-5}$ & 2.5 & ${\sim}1$-$2$ & $10^{-4}$ \\
Meinecke et al. \citep{M15} & Vulcan A & 0.2-0.5 & 3-7 & ${\sim}10^{-5}$ & $\lesssim 1$ &  ${\sim}1$-$3$ & $2 \times 10^{-7}$ \\
White et al. \citep{W19} & Vulcan B & 0.2 & $\lesssim 1$ & ${\sim}10^{-5}$ & $1$--$6$ & $<1$ & $2 \times 10^{-3}$ \\
Tzeferacos et al. \citep{T18} & OMEGA A & 0.06 & $\lesssim 600$ & 0.2-0.5 & $0.5$ & $\sim 30$ & 0.04  \\
Bott et al. \citep{B20} & OMEGA B & 0.04 & $\lesssim 450$ & 1-3 & $0.5$ & $\sim 20$ & 0.03 \\
Meinecke et al. \citep{M20} & NIF & 0.06 & $2000$--$6000$ & $3$--$50$ & $0.6$ & $\sim 13$ & 0.09 \\
\end{tabular}}
\caption{Plasma parameters attained in experiments that have investigated turbulent amplification of magnetic fields in laser plasmas. Here, $L$ is the driving scale of turbulent motions in the plasma; we note that in~\citep{M15}, the reported magnetic Reynolds number is instead defined with the scale length of the entire turbulent plasma. }
\label{table:ExpRef}
\end{table}
The parameters reported here have all been calculated using consistent definitions of relevant quantities (e.g., the length scales used in the fluid and magnetic Reynolds numbers); we caution readers that the conventions (and notation) employed in the original papers are not all consistent with each other. For convenience, we refer to each experiment via an identifier: the name of the laser facility at which the experiment was carried out. 

While amplification of magnetic fields is seen in all of the experiments, it is much more significant in the NIF, OMEGA A and OMEGA B experiments than in our LMJ experiment, or in the Vulcan experiment. The obvious difference between these two sets of experiments is the values of $\mathrm{Rm}$ and $\mathrm{Pm}$. More specifically, for the LMJ and Vulcan experiments, $\mathrm{Rm}$ is likely to be below the critical value $\mathrm{Rm}_c$ required for the turbulent dynamo to operate in the $\mathrm{Pm} \ll 1$ regime. In the OMEGA and NIF experiments, that critical threshold seems to have been successfully surpassed (indeed, in OMEGA B and NIF we anticipate that the greater than order-unity value of $\mathrm{Pm}$ reduces $\mathrm{Rm}_c$ significantly). 

\begin{figure}
  \centering
\includegraphics[width=0.95\linewidth]{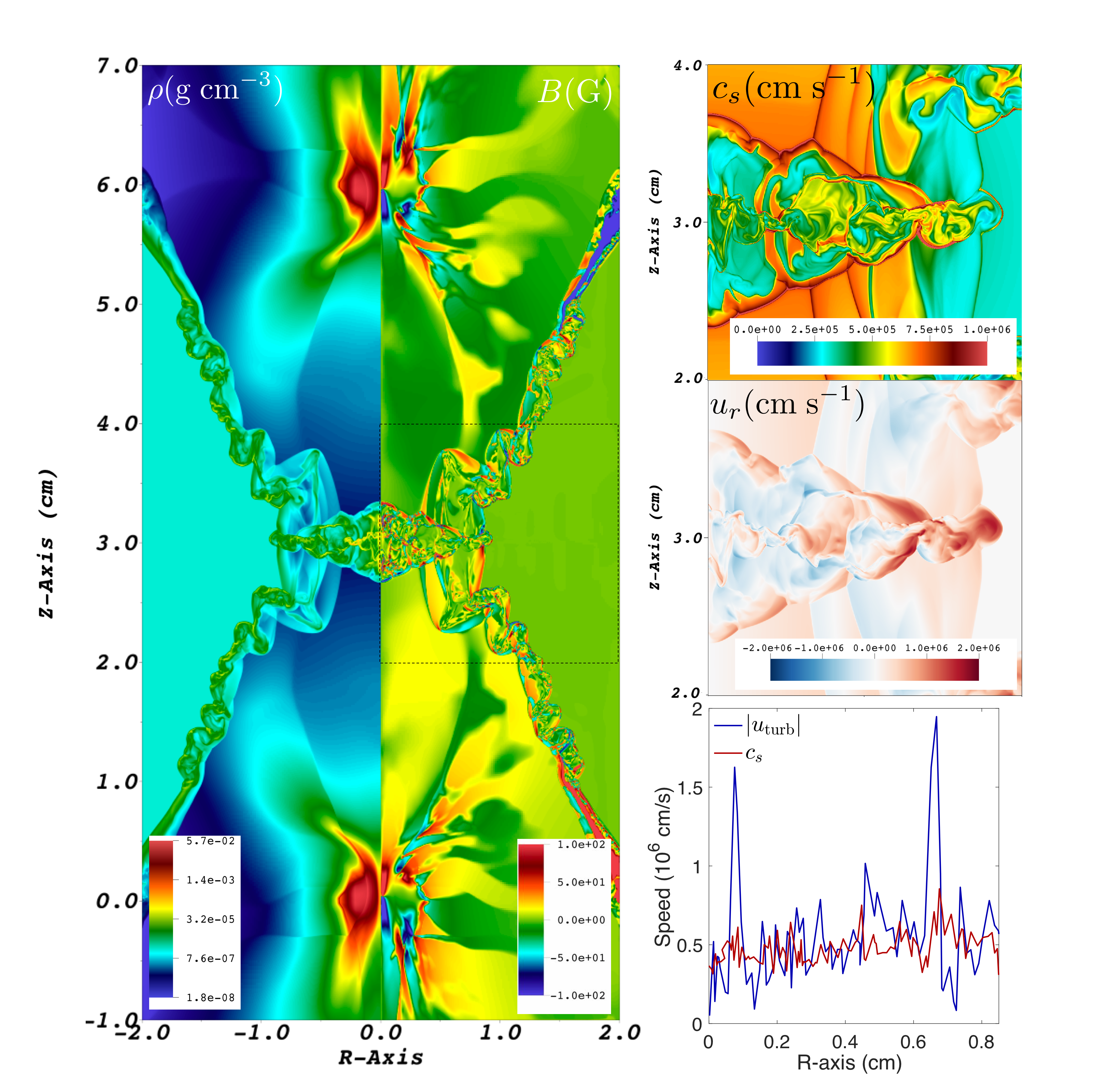}
\caption{\textit{Establishing turbulent Mach number of Vulcan experiment using FLASH simulations}. The simulations are two dimensional, and cylindrically symmetric (full details are provided in~\citep{M15}). Left panel: slice plots of density and magnetic field strength at time ($t = 950$ ns) at which peak field strengths are attained. Top right panel: sound speed $c_s$ calculated using the formula given in Table \ref{table:TableTheoParam} at the same time; the plasma is a carbon-argon mixture whose composition varies with position, so the simulated values of $M$ and $\langle Z \rangle$ are determined from the simulation. Middle right panel: radial component of velocity. Bottom right panel: lineout (taken at $Z = 3.0$ cm) of the sound speed and the total turbulent velocity (including both radial and axial components).}
\label{fig:FLASH_Vulcan}
\end{figure}

The similar degree of amplification seen in the LMJ experiment and in the Vulcan A experiment, in spite of the significantly smaller values of $\mathrm{Rm}$ obtained in the latter, suggests that another factor must be affecting the efficacy of the amplification mechanism in our experiment. As discussed in the main text, a plausible candidate is the turbulent Mach number. The Vulcan B experiment suggests that the turbulent Mach number of the plasma increases significantly with time~\citep{W19}; when comparing to our experiment, it is therefore necessary to determine the turbulent Mach number attained in the Vulcan experiment at comparable times to the peak field amplification (at $t = 950$ ns). Since this number is not reported in~\citep{M15}, we make this determination using validated FLASH MHD simulations of the Vulcan experiment. Figure \ref{fig:FLASH_Vulcan} shows slice plots of the density and magnetic field at $t = 950$ ns, as well as the sound speed and (one component of) the velocity in the interaction-region plasma. 
Taking a lineout across the interaction-region plasma, and comparing the mean turbulent velocity and mean sound speed, we find that, on average, $\mathrm{Ma}_{\rm turb} = u_{\rm turb}/c_s \approx 1$. We conclude that amplification in the Vulcan A experiment does not occur in the supersonic regime, whereas it does in the LMJ experiment. This difference likely explains why the LMJ experiment saw similar field amplification to the Vulcan A experiment, in spite of much greater characteristic values of $\mathrm{Rm}$ in the former.